\documentclass[numbers,authoryear]{article}

\usepackage[T1]{fontenc}
\usepackage[utf8]{inputenc}
\usepackage{fancyhdr}
\usepackage{setspace}
\usepackage[english]{babel}
\usepackage{a4wide}
\usepackage[pdftex]{graphicx}
\usepackage{pdfpages}
\usepackage[top=3cm, bottom=3cm, left=3cm, right=3cm]{geometry}
\usepackage{comment}
\usepackage{changepage}
\usepackage{graphicx}
\usepackage{paralist}
\usepackage{enumitem}
\usepackage{listings}
\usepackage{nth}
\usepackage{algorithm}
\usepackage{algpseudocode}
\usepackage{amsmath, amsthm, mathtools, amsfonts,mathrsfs, upgreek, mathabx}
\usepackage{adjustbox}
\usepackage{setspace}
\usepackage{here}
\usepackage{rotating}
\usepackage{multirow}
\usepackage{booktabs}
\usepackage{lettrine}
\usepackage{dsfont}
\usepackage{textcomp}
\usepackage[tikz]{bclogo}
\usepackage{fancybox}
\usepackage{pifont}
\usepackage{color, xcolor}
\usepackage{bbm}
\usepackage{multicol}
\usepackage{footnote}
\usepackage{supertabular, longtable, caption, afterpage, tabularx, threeparttable}
\usepackage{subcaption}
\usepackage{colortbl}
\usepackage{dcolumn}
\newcolumntype{d}[1]{D..{#1}} 
\usepackage{lscape}

\usepackage{authblk}
\usepackage[authoryear]{natbib}
\usepackage[colorlinks=true,urlcolor= blue,citecolor=blue]{hyperref}
\usepackage{titlesec}
\titleformat{\paragraph}
{\normalfont\normalsize\bfseries}{\theparagraph}{1em}{}
\titlespacing*{\paragraph}
{0pt}{3.25ex plus 1ex minus .2ex}{1pt}

\makeatletter
\newcommand{\neutralize}[1]{\expandafter\let\csname c@#1\endcsname\count@}
\makeatother
\newtheorem{assumption}{Assumption}[section]

\usepackage{ragged2e}
\title{\vspace{-2.5cm}\textbf{The Role of Child Gender in the Formation of Parents' Social Networks}\thanks{We thank Vincent Boucher for valuable comments. Asad Islam would like to acknowledge the funding support from FCDO/ESRC grant (grant number: ES/N010221/1: Investing in our Future: The Early Childhood Intervention and parental involvement in Bangladesh).}\\
}
\author{Aristide Houndetoungan\thanks{Department of Economics, Thema, Cy Cergy Paris Universit\'{e}. Email: aristide.houndetoungan@cyu.fr.} \ Asad Islam\thanks{Centre for Development Economics and Sustainability (CDES) and Department of Economics, Monash University, and  J-PAL.  Email: asadul.islam@monash.edu.}   \ Michael Vlassopoulos\thanks{Economics Department, Social Sciences, University of Southampton, and IZA. Email: m.vlassopoulos@soton.ac.uk} \ Yves Zenou\thanks{Department of Economics, Monash University,  CEPR, and IZA. Email: yves.zenou@monash.edu.}
}

\date{\today}

\definecolor{colm}{rgb}{0.45, 0.5, 0.9}
\definecolor{cola}{rgb}{0.7, 0, 0.7}
\definecolor{coly}{rgb}{0.2, 0.7, 0.5}

\begin{document}

\maketitle
\begin{abstract}
\noindent Social networks play an important role in various aspects of life. While extensive research has explored factors such as  gender, race, and education in network formation, one dimension that has received less attention is the gender of one's child. Children tend to form friendships with same-gender peers, potentially leading their parents to interact based on their child's gender. Focusing on households with children aged 3-5, we leverage a rich dataset from rural Bangladesh  to investigate the role of children's gender in parental network formation. We estimate an equilibrium model of network formation that considers a child's gender alongside other socioeconomic factors. Counterfactual analyses reveal that children's gender significantly shapes parents' network structure. Specifically, if all children share the same gender, households would have approximately 15\% more links, with a stronger effect for families having girls. Importantly, the impact of children's gender on network structure is on par with or even surpasses that of factors such as income distribution, parental occupation, education, and age. These findings carry implications for debates surrounding coed versus single-sex schools, as well as policies that  foster inter-gender social interactions among children.
\end{abstract}

\vspace{0.5cm}



\newpage
\section{Introduction}

Social networks play a fundamental role in shaping various behaviors and influencing multiple aspects of our lives, from educational choices, to job opportunities, and access to information \citep{jackson2008social,jackson2017economic,jackson2019}. Previous research has extensively explored the formation of social networks, emphasizing factors such as gender, race, ethnicity, and social and geographic proximity (homophily) as key determinants of network connections \citep{mcpherson2001birds,currarini2009economic,boucher2015structural,fafchamps2007formation}. However, one dimension that has received relatively less attention, despite its potential significance, is the gender of children within households. Indeed, during early childhood, girls and boys spend much of their time at home with their families and seek guidance from parents. In this paper, we delve deeper into this relatively unexplored dimension of parental networks.


Why might the gender of a child be an important determinant of parents' social links? Several underlying mechanisms are plausible. Firstly, parents may actively seek opportunities to connect with other parents who have children of the same gender. Secondly, children of the same gender tend to engage in more frequent social interactions \citep{maccoby1987gender}, which could naturally lead to even more interactions among their parents, as they come together for playdates, school and social events, and various extracurricular activities.\footnote{In the psychology literature, when  researchers observed correlations between parenting practices and children’s behavior, traditionally they inferred  that  parents were influencing their children; however, developmental psychologists now recognize that children also influence their parents’ behavior, particularly in the context of parental socialization \citep{Leaper2014}.} These connections can serve as valuable sources of information and support, particularly in matters related to child rearing. For instance, parents may exchange insights on gender-specific upbringing challenges and share strategies for addressing them. This gender-specific information exchange can further strengthen social ties among parents. A better understanding of the role of children's gender in parental network formation is important as these parental connections have the potential to impact children's access to resources, information, social networks, and opportunities, making it a critical component of children's human capital development \citep{Guo2023}.


We leverage a rich dataset of household networks across 222 villages in rural Bangladesh, encompassing about 4,500 households with children aged 3-5. The network data are based on a detailed pairwise network census of participating households, capturing various dimensions of their interactions. This dataset provides a unique opportunity to investigate the role of children's gender in parental social network formation in a context where social networks play an important role and can act as substitutes for formal institutions, for example, in the provision of insurance against income or health shocks \citep{fafchamps2003risk,de2006risk,breza2016field,breza2019networks}. We estimate a version of the equilibrium model of network formation developed by \cite{mele2017structural,mele2020does,mele2022structural} in which individuals may receive distinct utility benefits from forming direct, mutual, and indirect links. In estimating the model, we allow the influence of a child's gender on the likelihood of parents forming connections, alongside other socioeconomic factors known to influence the formation of network links, such as occupation and education.  

Our structural estimates reveal that children's gender is an important determinant of parents' networks. Moreover, we find evidence of indirect homophily by children's gender, that is, the influence of children's gender extends beyond direct links: families prefer to form links with other families that are in turn connected to families with same-gender children. We then use the model to perform counterfactual simulations to gain insight on how changing the gender composition of children affects the structure of the network. This analysis reveals that children's gender is an important factor for the formation of network links. Our simulations show that if all children have the same gender, then households would have about 15\% more links. Interestingly, the effect of gender on link formation of families is stronger for pairs of families that have girls rather than boys. We also examine, through policy experiments, the importance of other factors for parents' network structure, such as income inequality, parents' occupation, education, and age. Remarkably, we find that, in comparison to children's gender, most of these other key factors carry similar or even less quantitative importance in shaping parents' networks.

This paper contributes to the empirical literature on network formation (for overviews, see \cite{Graham2015}, \cite{chandrasekhar2016econometrics}, and \cite{Paula2017, de2020econometric}) by providing novel insights into the so far overlooked role of children's gender within parental networks. Methodologically, the paper most closely related to ours is \cite{mele2020does}, which estimates the network formation framework developed in \cite{mele2017structural} to study racial segregation in schools and assess how student reallocation policies impact segregation through simulations. We take advantage of our rich dataset to address the non-identification issue in exponential random graphs pointed out by \cite{mele2017structural}. As we observe networks across a large number of villages, we can leverage variation in network features across these villages to identify the model parameters. Other notable contributions in this literature include works by \cite{currarini2009economic}, \cite{leung2015two}, \cite{boucher2015structural}, \cite{boucher2017my}, \cite{graham2017econometric}, \cite{christakis2020empirical}, \cite{BTVWZ21}, \cite{mele2022structural}, \cite{hsieh2022structural}, and \cite{boucher2023education}. The majority of these studies emphasize factors such as gender, race, ethnicity, and social and geographic proximity as key determinants of network connections. To the best of our knowledge, our paper is the first to explicitly consider the role of children's gender within parental networks.\footnote{\cite{Heizler2017} study the effect of family composition (e.g., number of children and 
 age gap between the oldest and youngest child) on parents' social interactions. The analysis is very different since they do not have network data and thus, do not model the network formation of parents and the role of children's gender on their parents' networks. There is also a literature on the interaction between parents and children in terms of education (for an overview, see \cite{Doepke2019}) but the focus is very different since they look at how various parenting styles affect children's education outcomes.}  

Our results speak to the ongoing debate around coed (coeducational) schools versus single-sex (same-sex) schools. A large literature has documented that a single-sex environment might facilitate more effective learning for females; this is particularly true at a younger age \citep{Sullivan2010}. Specifically, reduced gender stereotype threat might lead to increased self-confidence \citep{Spencer1999, Booth2014, Eisenkopf2015} and, in the absence of male peers, females might also feel less anxiety pursuing non-stereotypical courses, such as math \citep{Mael2004, Hahn2020}. Our counterfactual exercises suggest that single-sex schools could lead to parental networks that are more dense, especially in the case of girls, potentially influencing parents' decisions regarding their children's education. Beyond schools, opportunities for social interactions among children arise in participation to youth clubs and organizations, and various religious and community events. Our findings indicate that promoting social interactions among children, irrespective of their gender, can have a meaningful impact on parental networks, potentially influencing the flows of information, social support, and cohesion in these communities, which might even feedback into the upbringing of the children.

The rest of the paper is organized as follows. In Section \ref{sec:context}, we introduce the context and present the dataset, offering descriptive evidence that highlights the influence of children's gender on parental networks. Section \ref{sec:model} presents a dynamic network model in which each parent's network decisions are influenced by observable characteristics and the networks of other parents. The estimation results are presented in Section \ref{sec:result}. Additionally, we present counterfactual analyses in this section to examine how specific characteristics can impact the network. Finally, Section \ref{sec:conclusion} offers some concluding remarks.

\section{Context, Data, and Descriptive Evidence\label{sec:context}}

Our data collection was conducted in 222 villages located in two rural districts of Bangladesh, Khulna and Shatkhira. These villages had been selected to be part of an early childhood intervention targeting 3-5 year old children over a 2-year period (see \cite{Guo2023} for details). In each of these participating villages, we recruited all eligible families, defined as those having at least one child in the 3-5 year old age group, residing in the vicinity where an early childhood center was planned to be established for the intervention. Consequently, we recruited an average of 20 households in each village, with numbers ranging from 7 to 37, resulting in a total sample of 4,473 families. The parents' social network information employed in the current study was collected at the baseline from April to May 2017, preceding the implementation of the intervention. During this data collection, we gathered a comprehensive array of information from surveyed households concerning their education, occupation, land ownership, income, expenditures and savings, and child-rearing practices. 

Furthermore, we asked several questions that allow us to discern the social ties among all surveyed households within each group of eligible households. Specifically, the survey included questions on the extent to which each pair of families interact with each other. This included whether each household borrows from each of the other households in the same group and whether they seek help from each other when a family member is sick. We make use of these questions to define network links among families, following a similar approach as in \cite{banerjee2013diffusion,banerjee2021changes}. Given the rural context of these communities, it is reasonable to expect that families relying on each other  for risk-sharing and support would have close-knit social bonds. Therefore, our definition of a link is designed to capture these meaningful and important relationships within the community.

\paragraph{Defining links and networks}
The focus of our study is the formation of social networks among parents within these villages. To be more precise, the network is defined among the group of eligible households that were surveyed within each village. Hereafter, we use the term ``village networks'' to refer to the networks established within this group of households in each village. Formally, consider any pair of surveyed parents, denoted by $i$ and $j$, belonging to the set of eligible parents $\mathcal{V}_r$ in village $r$. We define a binary variable, $a_{r,ij}$, such that $a_{r,ij} = 1$ if parent $i$ lists parent $j$ as a friend and $a_{r,ij} = 0$ otherwise. Family $j$ is connected to family $i$ if at least one of the following conditions holds: $i$ can ask for help from $j$ when $i$ needs to borrow 100 Taka; $i$ can ask for help from $j$ when someone in family $i$ falls ill. 

It is important to note that we observe the entire network among households within each group $\mathcal{V}_r$. Furthermore, households are not considered to be connected to themselves, that is, $a_{r,ii} = 0$ for any $r$ and $i$. The network is directed, meaning that if $j$ is listed as a connection by $i$, it does not imply that $i$ is a connection to $j$. This configuration matches our dataset and is consistent with existing literature \citep[see][]{graham2020econometric}. We also restrict interactions to the same village: household $i$ can only seek help in their own village and not in neighboring ones. Finally, the social network is independent across villages. Thus, the network in the  village $r$ can be represented by an adjacency matrix $A_r = [a_{r,ij}]_{\substack{1\leq i \leq n_r\\ 1\leq j \leq n_r}} \in \mathcal{A}_r$, where $\mathcal{A}_r$ is the set of possible directed adjacency matrices in village $r$.

\paragraph{Descriptives of networks and households}
Making use of these data, we construct individual networks for each village at the household level.\footnote{See \cite{banerjee2013diffusion} and \cite{breza2019social} for a related approach to defining a network among village households.} Summary statistics for both the networks and the households are presented in Table \ref{tab:data}. The average network size (number of network members) is approximately 20.15 households, with a range from 7 to 37 households. The average degree (number of links) is 6.56, with the number ranging from 0 to 28 for some households. The high clustering coefficient of 0.72 indicates a propensity for triads to form clusters within these networks. 

\begin{figure}[htbp]
    \centering
    \includegraphics[scale = 0.75]{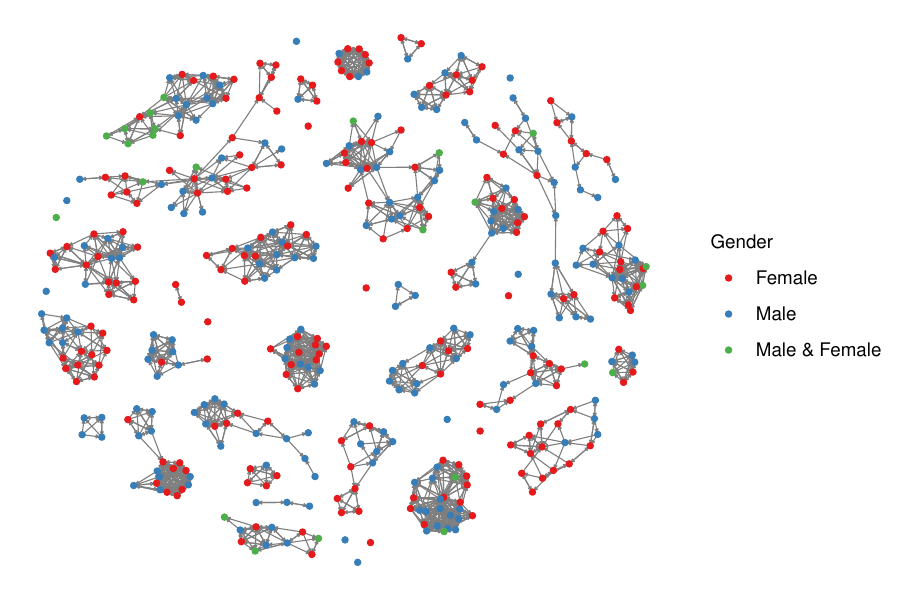}
    \caption{Networks of Selected Villages}
    \label{fig:net}
\end{figure}

Figure \ref{fig:net} illustrates the networks within a selected 10\% subset of villages from our sample. Each node in this graph corresponds to a family, with families having girls represented by red nodes, those having boys represented by blue nodes, and those with both girls and boys depicted as green nodes. It is noteworthy that nodes of the same color tend to cluster together in the graph, indicating that families with children of the same gender tend to form connections within the network. Notably, the formation of these clusters appears to be more prominent among families with girls.

Turning to household characteristics, children are balanced in terms of gender. Given the presence of young children in these households, fathers have an average age of about 34 years, while mothers average around 27 years. In terms of education, a significant portion of fathers, roughly 73\%, have completed primary or secondary education, while 85\% of mothers have attained similar levels of education. In terms of occupation, fathers in the sample are mostly daily laborers, self-employed, or farmers. In contrast, the majority of mothers in this sample do not work outside of the household.

\begin{table}[!htbp]
  \centering 
  \small
  \caption{Data summary} 
  \label{tab:data} 
\begin{threeparttable}
\begin{tabular}{lrrrr}
\toprule
                               & \multicolumn{1}{c}{Mean}   & \multicolumn{1}{c}{SD}        & \multicolumn{1}{c}{Min}  & \multicolumn{1}{c}{Max}   \\ \midrule
\textbf{Network}                &        &        &      &        \\
Degree                         & 6.56   & 4.41   & 0    & 28     \\
Density                        & 0.31   &    & &   \\
Clustering                     & 0.72   &    &  &    \\
Asymmetry                      & 0.22   &    \\
Group's Size                   & 20.15  & 5.42   & 7    & 37     \\
                               &        &        &      &        \\
\textbf{Household}                      &        &        &      &        \\Male children                  & 0.50   & 0.50   & 0   & 1      \\
Female children                & 0.50   & 0.50   & 0   & 1      \\
Father’s age                   & 34.19  & 6.12   & 20  & 82     \\
Mother’s age                   & 27.18  & 5.00   & 20  & 55     \\
Income ($10^3$ Taka)           & 11.20 & 5.06 & 0    & 66 \\
Homestead land ($10^3$ Ha)     & 0.084   & 0.114  & 0   & 0.230    \\
Farming land ($10^3$ Ha)       & 0.317  & 1.700 & 0   & 100 \\
Father’s education: none       & 0.16   & 0.37   & 0   & 1      \\
Father’s education: primary    & 0.32   & 0.47   & 0   & 1      \\
Father’s education: secondary  & 0.41   & 0.49   & 0   & 1      \\
Father’s education: university & 0.11   & 0.31   & 0   & 1      \\
Mother’s education: none       & 0.08   & 0.27   & 0   & 1      \\
Mother’s education: primary    & 0.23   & 0.42   & 0   & 1      \\
Mother’s education: secondary  & 0.62   & 0.49   & 0   & 1      \\
Mother’s education: university & 0.08   & 0.27   & 0   & 1      \\
Daily laborer fathers          & 0.35   & 0.48   & 0   & 1      \\
Farmer fathers                 & 0.21   & 0.41   & 0   & 1      \\
Self-employed fathers           & 0.31   & 0.46   & 0   & 1      \\
Mother without a job           & 0.93   & 0.26   & 0   & 1      \\
\bottomrule
\end{tabular} 
\begin{tablenotes}[para,flushleft]
\end{tablenotes}
\end{threeparttable}
\end{table}

\paragraph{Descriptives of link formation} To gain an initial, descriptive insight into the factors influencing the formation of links between households, Table \ref{TAB:logit} presents results of logit regressions. These regressions estimate the probability of a link between household $i$ and $j$ based on various household characteristics, including whether the children in the two households share the same gender and various other socio-economics characteristics of the households. Specifically, we estimate two models defined as follows: 
\begin{equation}\label{eq:log}
\frac{\mathbb{P}(a_{r,ij} = 1)}{1 - \mathbb{P}(a_{r,ij} = 1)} =
 \begin{cases}
     \exp(\theta_1 Gender_{r,ij} + \boldsymbol{x}_{r,ij}^{\prime}\boldsymbol{\theta}_2),  \\
     \exp(\theta_1^m Male_{r,ij} + \theta_1^f Female_{r,ij} +\boldsymbol{x}_{r,ij}^{\prime}\boldsymbol{\theta}_2) , \\
 \end{cases}
\end{equation}
\noindent where  $Gender_{r,ij}$,  $Male_{r,ij}$, and $Female_{r,ij}$ are dummy variables, such that
$Gender_{r,ij} = 1$ if $i$ and $j$ have children of same gender, $Male_{r,ij} = 1$ if they both have boys and $Female_{r,ij} = 1$ if they have girls. The second equation in \eqref{eq:log} is a more flexible version of the first one. The first specification in \eqref{eq:log} imposes that $\theta_1^m = \theta_1^f = \theta_1$. The variable $\boldsymbol{x}_{r,ij}$ is a vector of control factors susceptible to influence the network, and $\theta_1$, $\theta_1^m$, $\theta_1^f$, and $\boldsymbol{\theta}_2$ are unknown parameters to be estimated.

In Column (1), we estimate a single parameter for the effect of same gender corresponding to the first specification in Equation \eqref{eq:log}, whereas in Column (2) (second specification in Equation \eqref{eq:log}) we allow the gender of children to have different effects. In Column (1), we find that having same gender children has a positive effect on the probability of a link, but the effect is not statistically significant. In Column (2), we observe gender differences, with having boys in both households having a positive and statistically significant effect on link formation, while having girls has a positive but non-significant effect. Beyond the gender of the children, these regressions highlight that households with similar characteristics, such as age, education, occupation, income, and land ownership of parents, are more likely to be linked. These results provide a preliminary understanding of the factors influencing social network formation among parents in the studied villages.

\begin{table}[!htbp]
  \centering 
  \small
  \caption{Standard logit model for network formation} 
  \label{TAB:logit} 
\begin{threeparttable}
\begin{tabular}{@{}ld{7}ld{7}l@{}}
\toprule
                                      & \multicolumn{2}{c}{(1)} & \multicolumn{2}{c}{(2)} \\
                          \midrule
Intercept                             & -0.998^{***} & (0.069) & -1.004^{***} & (0.070) \\
$\text{Male}_{ij}$                    &                                 &         & 0.043^{***}  & (0.017) \\
$\text{Female}_{ij}$                  &                                 &         & 0.012     & (0.017) \\
$\text{Gender}_{ij}$                  & 0.021     & (0.014) &                                 &         \\
Adiff(Father’s Age)                   & 0.002     & (0.002) & 0.002     & (0.002) \\
$\text{Age Father}_i$                 & -0.002    & (0.002) & -0.002    & (0.002) \\
$\text{Age Father}_j$                 & -0.005^{**}  & (0.002) & -0.005^{**}  & (0.002) \\
Adiff(Mother’s Age)                   & -0.005^{**}  & (0.002) & -0.005^{**}  & (0.002) \\
$\text{Age Mother}_i$                 & 0.002     & (0.002) & 0.002     & (0.002) \\
$\text{Age Mother}_j$                 & 0.006^{**}   & (0.002) & 0.006^{**}   & (0.002) \\
Adiff(Income)                         & -0.016^{***} & (0.002) & -0.016^{***} & (0.002) \\
$\text{income}_i$                     & 0.012^{***}  & (0.002) & 0.012^{***}  & (0.002) \\
$\text{Income}_j$                     & 0.013^{***}  & (0.002) & 0.013^{***}  & (0.002) \\
Adiff(Homestead land)                 & -0.411^{***} & (0.119) & -0.412^{***} & (0.119) \\
$\text{Homestead land}_i$             & 0.300^{***}    & (0.112) & 0.300^{***}    & (0.112) \\
$\text{Homestead land}_j$             & 0.191^{*}    & (0.112) & 0.192^{*}    & (0.112) \\
Adiff(Farming land)                   & -0.185^{***} & (0.017) & -0.185^{***} & (0.017) \\
$\text{Farming land}_i$               & 0.179^{***}  & (0.017) & 0.179^{***}  & (0.017) \\
$\text{Farming land}_j$               & 0.189^{***}  & (0.017) & 0.189^{***}  & (0.017) \\
$\text{Father’s education}_{ij}$      & 0.055^{***}  & (0.015) & 0.055^{***}  & (0.015) \\
$\text{Mother’s education}_{ij}$      & 0.031^{**}   & (0.015) & 0.031^{**}   & (0.015) \\
$\text{Father is daily laborer}_{ij}$ & 0.268^{***}  & (0.021) & 0.268^{***}  & (0.021) \\
$\text{Father is farmer}_{ij}$        & 0.151^{***}  & (0.029) & 0.152^{***}  & (0.029) \\
$\text{Father is self-employed}_{ij}$ & -0.098^{***} & (0.023) & -0.098^{***} & (0.023) \\
$\text{Mother stays home}_{ij}$       & -0.073^{***} & (0.021) & -0.072^{***} & (0.021)\\
\bottomrule
\end{tabular} 
\begin{tablenotes}[para,flushleft]
\footnotesize
Note: ***: \textit{pvalue $<1\%$}, **: \textit{pvalue $<5\%$}, *: \textit{pvalue $<10\%$}\\
This table presents estimation results on link formation using a logit model. The first column for each model indicates the coefficient estimate followed by their standard error in brackets. For any variable denoted "var", "var$_{ij}$" takes one if var$_i =$ var$_j$ and zero otherwise; ``Adiff(var)" if the absolute difference between var$_i$ and var$_j$. The network is directed and the relationship we model is the link from $j$ to $i$. There are 4,473 families, and 92,152 potential links in the sample.
\end{tablenotes}
\end{threeparttable}
\end{table}

The aforementioned results impose that $(a_{r,ij}, ~Gender_{r,ij}, ~Male_{r,ij}, ~Female_{r,ij}, ~\boldsymbol{x}_{r,ij})$ are independent across $r$, $i$, and $j$. This assumption implies that family $i$ does not consider the potential connections between $j$ and other families when deciding to form a link with $j$. Given the definition of the network used in this paper, this assumption may be overly restrictive. For instance, family $i$ is more likely to interact with family $j$ if the latter has access to resources or has friends with such access. Therefore, we need to consider a more flexible model to account for these complexities.

\section{A Model of Parent Network Formation\label{sec:model}}
This section presents the model and the estimation strategy. We adapt the framework of \cite{mele2017structural}  to our specific context in which we have many villages (groups) ($\Bar{r}\in\mathbb{N}^{\ast}$). From a theoretical point of view, the model remains the same because we assume that the network is independent across villages \cite[see][]{mele2022structural}. Let $\mathcal{V}_r = \{1, ~\dots,n_r\}$ denote a set of $n_r$ parents (families) within village $r$, for $r = 1,~\dots,~\Bar{r}$. Each family $i\in \mathcal{V}_r$ is characterized by a vector of observed characteristics assumed to be exogenous and denoted by $\boldsymbol{x}_{r,i}$. This vector $\boldsymbol{x}_{r,i}$ encompasses various attributes such as the presence of a male child, parents' ages, father's occupation, family income, etc.

\subsection{Microeconomic foundations}
\paragraph{Family preferences}
The model is dynamic and time is discrete.  The realization of the network at time $t$ is denoted by $A_r^t = [a_{r,ij}^t]_{\substack{1\leq i \leq n_r\\ 1\leq j \leq n_r}}$, where $a_{r,ij}^t = 1$ if parent $j$ is connected with  parent $i$  at time $t$ and $a_{r,ij}^t = 0$ otherwise. In each village $r$, at the beginning of each period, a family $i$ is selected at random to meet family $j$ according to a meeting probability conditional on the set of family characteristics $\mathcal{X}_r := \{\boldsymbol{x}_{r,i}, ~\dots, ~\boldsymbol{x}_{r,n_r}\}$ and the realization of the network at time $t - 1$. For instance, family $i$ is likely to meet family $j$ if they have common friends at time $t-1$ or if they share similar characteristics such as being both farmers or having attended the same school. 

It is important to note that a meeting does not necessarily result in the formation of a link. Family $i$ can meet family $j$, but if the interaction doesn't result in a strong connection, $i$ may not consider $j$ as someone they would ask for help. Upon meeting, family $i$ decides whether to update their link $a_{r,ij}^t$. In this model, information is complete, in the sense that families observe the entire network (in its dynamic form) and the set of family characteristics $\mathcal{X}_r$ before making their choice regarding forming a link. When family $i$ selects their friends, they derive utility given by

\begin{equation}
    U_{r,i}(A_r, \mathcal{X}_r) = \underbrace{\sum_{\substack{j = 1 \\ j \ne i}}^{n_r} a_{r,ij}u_{r,ij}}_{\text{direct friends}} + \underbrace{\sum_{\substack{j = 1 \\ j \ne i}}^{n_r} a_{r,ij}a_{r,ji}m_{r,ij}}_{\text{mutual friends}} + \underbrace{\sum_{\substack{j = 1 \\ j \ne i}}^{n_r}\sum_{\substack{k = 1 \\ k \ne i, j}}^{n_r} a_{r,ij}a_{r,jk}v_{r,ik}}_{\text{friends of friends}} + \underbrace{\sum_{\substack{j = 1 \\ j \ne i}}^{n_r}\sum_{\substack{k = 1 \\ k \ne i, j}}^{n_r} a_{r,ij}a_{r,ki}w_{r,kj}}_{\text{popularity}},\label{eq:utility}
\end{equation}
\noindent where $u_{r,ij} \equiv u(\boldsymbol{x}_{r,i}, \boldsymbol{x}_{r,j})$, $m_{r,ij} \equiv m(\boldsymbol{x}_{r,i}, \boldsymbol{x}_{r,j})$, $v_{r,ik} \equiv v(\boldsymbol{x}_{r,i}, \boldsymbol{x}_{r,k})$, and $w_{r,kj} \equiv w(\boldsymbol{x}_{r,k}, \boldsymbol{x}_{r,j})$ are bounded real-valued functions of the observed family characteristics $\boldsymbol{x}_{r,i}$,  $\boldsymbol{x}_{r,j}$, and $\boldsymbol{x}_{r,k}$.\footnote{We omit the superscript $t$ to ease the notational burden.} 

Utility function \eqref{eq:utility} is quite general and accounts for multiple factors that can influence the network formation process. The first term captures the utility derived from direct friendships. Family $i$ receives a utility of $u_{r,ij}$ if they form a link with family $j$. Given that $u_{r,ij}$ depends on both $\boldsymbol{x}_{r,i}$ and $\boldsymbol{x}_{r,j}$, then $u_{r,ij}$ can be defined as a distance between $\boldsymbol{x}_{r,i}$ and $\boldsymbol{x}_{r,j}$; that is, a measure of social distance between families $i$ and $j$. Consequently, \eqref{eq:utility} serves as a generalized framework that can encompass specific cases where network formation is driven by homophily \citep{mcpherson2001birds}. 

The second term in \eqref{eq:utility} represents the additional utility ($m_{r,ij}$) derived from friends who reciprocate the connection by also considering family $i$ as a friend (mutual relationships). This term holds particular significance within our framework for several reasons. Firstly, it addresses the reciprocity of assistance. When family $i$ seeks help, their friend $j$ may be hesitant to provide aid. However, if $j$ is aware that they can also turn to $i$ for assistance in the future, they are more likely to agree to provide assistance. On the other hand, because of the assumption of perfect information, family $i$ will not hesitate to request help if they are aware that their friend $j$ can also seek help from them. Therefore, mutual relationships offer family $i$ a distinct benefit compared to direct friendships without such mutual connections.

The third term in \eqref{eq:utility} represents the additional utility derived from indirect relationships (friends of friends). In this term, family $i$ accrues a utility benefit $v_{r,ik}$ for every pair of families $j$ and $k$ where $j$ is a direct friend of $i$ and $k$ is a friend of $j$, irrespective of whether $k$ is a direct friend of $i$ or not. For simplicity, we assume, as in \cite{mele2017structural}, that $v_{r,ik}$ depends only on the observable characteristics of family $i$ and family $k$,  but not those of intermediary family $j$. In our context, these indirect relationships are relevant because they facilitate the potential for  family $i$ to receive assistance from family $k$ through the intermediary connection of family $j$, even if family $k$ is not a directly connected to family $i$.

Finally, the last term in \eqref{eq:utility} accounts for popularity within the network. In this term, family $i$ receives a utility benefit $w_{r,kj}$ for every pair of families $j$ and $k$ where $j$ is a direct friend of $i$ and $i$ is a direct friend of $k$. Being popular can have both beneficial and costly implications for $i$. The utility term $w_{r,kj}$ serves to capture the combined impact of receiving assistance from family $j$ and the potential cost associated with providing support to family $k$. It reflects the intricate dynamics of network formation, where the benefits of popularity are counterbalanced by the obligations that come with maintaining connections.

\paragraph{A potential game}
The terms accounting for benefits arising from indirect relationships and popularity introduce a significant complication: it makes family $i$'s linking decision dependent on  decisions made by other families in the village. Although this feature of the model is relevant for our setting, it also poses significant challenges when attempting to estimate the econometric model without imposing additional constraints on preferences. The observational unit in the econometric model is the entire network at the village level (consisting of the group of surveyed households) rather than the entry $a_{r,ij}$, as is the case in classical dyadic models. This perspective  exponentially expands the number of potential network structures  as the number of possible network structures in village $r$ is $2^{(n_r - 1)n_r}$. For instance, in a village with just $n_r = 10$ families, the number of possible network configurations reaches a staggering $1.24\times10^{27}$, thereby rendering the estimation of the model computationally infeasible, even using supercomputers. To address this computational challenge, it is common practice to introduce restrictions for the game to have a potential function \citep[see][]{graham2020econometric}. 

\begin{assumption}[Preferences] For any $i$, $j$, $k$, the preferences satisfy the restrictions $m(\boldsymbol{x}_{r,i}, \boldsymbol{x}_{r,j}) = m(\boldsymbol{x}_{r,j}, \boldsymbol{x}_{r,i})$, and $v(\boldsymbol{x}_{r,k}, \boldsymbol{x}_{r,j}) = w(\boldsymbol{x}_{r,k}, \boldsymbol{x}_{r,j})$.\label{ass:preferences}
\end{assumption}
\noindent The first restriction is necessary for the identification of the model. If two families $i$ and $j$ are mutually linked, they should receive the same additional payoff in the second term of \eqref{eq:utility}. The second restriction requires that family $i$ values their popularity as much as family $k$ values the indirect relationship. This restriction ensures  the model's coherency in the sense of \cite{tamer2003incomplete}. Under Assumption \ref{ass:preferences}, \cite{mele2017structural} shows that the network formation process in each village is a potential game  \citep{monderer1996potential}, where a potential function in village $r$ is given by
\begin{equation}
    Q_r(A_r, \mathcal{X}_r) = \sum_{i = 1}^{n_r} \sum_{\substack{j = 1 \\ j \ne i}}^{n_r} a_{r,ij}u_{r,ij} + \sum_{i = 1}^{n_r} \sum_{\substack{j > i}}^{n_r} a_{r,ij}a_{r,ji}m_{r,ij} + \sum_{i = 1}^{n_r}\sum_{\substack{j = 1 \\ j \ne i}}^{n_r}\sum_{\substack{k = 1 \\ k \ne i, j}}^{n_r} a_{r,ij}a_{r,jk}v_{r,ik}.\label{eq:potential}
\end{equation}

\noindent The existence of a potential function greatly simplifies the characterization of families' incentives to form a link within the network. Consider two networks  $A_{r,ij}$ and $B_{r,ij}$, which  are identical in all respects except for a single entry $(i,~j)$. The difference in utility for family $i$ between these two networks, $U_{r,i}(A_{r,ij}, \mathcal{X}_r) - U_{r,i}(B_{r,ij}, \mathcal{X}_r)$, is equivalent to the difference in the potential function, $Q_r(A_{r,ij}, \mathcal{X}_r) - Q_r(B_{r,ij}, \mathcal{X}_r)$. This equivalence implies a straightforward criterion for family $i$: it prefers $A_{r,ij}$ over $B_{r,ij}$ if and only if $Q_r(A_{r,ij}, \mathcal{X}_r) > Q_r(B_{r,ij}, \mathcal{X}_r)$. Importantly, the function $Q_r$ is an aggregated measure of the entire network and is not specific to any particular family, as is the case with the utility function $U_{r,i}$. This fact has practical implications for empirical strategies. It allows us to sidestep the daunting computational challenge of analyzing $2^{(n_r - 1)n_r}$ distinct networks for each village.

\paragraph{Network distribution}
In addition to  utility $U_{r,ij}$, family $i$ receives an idiosyncratic shock $\varepsilon_{r,ij}(a_{r,ij})$ when they meet another family $j$. This shock is specific to the particular link configuration $a_{r,ij}$ between families $i$ and $j$ in village $r$. Based on their preferences at time $t$, family $i$ decides to establish a link with family $j$ if and only if it results in a higher utility, taking into account the impact of the shock. The idiosyncratic shock $\varepsilon_{r,ij}(a_{r,ij})$ is known to all families due to the assumption of complete information. However, the econometrician does not observe this shock, which makes the network at time $t$ a random variable due to the inherent uncertainty associated with these idiosyncratic shocks..

Although the network is dynamic, we only observe a single snapshot at a specific point in time. To estimate preferences using this limited observation of the network, \cite{mele2017structural} demonstrated that the dynamic network generated by the model followed a Markov Chain of networks that converges to a unique stationary distribution $\pi$ given by

\begin{equation}
    \pi(A_r|\mathcal{X}_r) = \dfrac{\exp(Q_r(A_r, \mathcal{X}_r))}{\sum_{W \in \mathcal{A}_r} \exp(Q_r(W, \mathcal{X}_r))}. \label{eq:pi}
\end{equation}

\noindent The denominator $\sum_{W \in \mathcal{A}_r} \exp(Q_r(W, \mathcal{X}_r))$  ensures that $\pi(.|\mathcal{X}_r)$ is a proper probability\textemdash   that is, it satisfies $\sum_{W \in \mathcal{A}_r}\pi(W|\mathcal{X}_r) = 1$ for any $r$. The notion of convergence to a stationary distribution implies that, for practical estimation purposes, the focus can be shifted towards estimating this stationary distribution $\pi$. This approach is underpinned by the assumption that the single network observation available to us originates from this distribution $\pi$. In essence, this perspective simplifies the estimation task, as it allows us to concentrate on understanding the network's long-term equilibrium behavior encapsulated by $\pi$.

\subsection{Econometric model and estimation strategy}
In this section, we present the econometric specification of the model and discuss its identification. We also outline a Bayesian approach for estimating the model parameters. 

\paragraph{Econometric specification}
Our objective is to estimate $u(\boldsymbol{x}_{r,i}, \boldsymbol{x}_{r,j})$, $m(\boldsymbol{x}_{r,i}, \boldsymbol{x}_{r,j})$, and $v(\boldsymbol{x}_{r,i}, \boldsymbol{x}_{r,k})$. This estimation would be sufficient to approximate the stationary distribution  $\pi$ in Equation \eqref{eq:pi} and quantify the impact of changes in $\boldsymbol{x}_{r,i}$ (e.g., change in the proportion of families with a boy) on the distribution $\pi$.
We assume that 
\begingroup
\begin{align}
\begin{split}\label{eq:spec}
    u(\boldsymbol{x}_{r,i}, \boldsymbol{x}_{r,j}) &= \beta^{u}_{1} Gender_{r,ij} + \boldsymbol{x}^{u\prime}_{ij}\boldsymbol{\beta}^{u}_{2},\\
    m(\boldsymbol{x}_{r,i}, \boldsymbol{x}_{r,j}) &= \beta^{m}_{1} Gender_{r,ij} + \boldsymbol{x}^{m\prime}_{ij}\boldsymbol{\beta}^{m}_{2},\\
    v(\boldsymbol{x}_{r,i}, \boldsymbol{x}_{r,k}) &= \beta^{v}_{1} Gender_{r,ik} + \boldsymbol{x}^{v\prime}_{ik}\boldsymbol{\beta}^{v}_{2},
\end{split}
\end{align}
\endgroup

\noindent where $Gender_{r,ij}$ is equal to 1 if families $i$ and $j$ have children of the same gender, $\boldsymbol{x}^{u}_{ij}$, $\boldsymbol{x}^{m}_{ij}$, and $\boldsymbol{x}^{v}_{ik}$ are vectors of control variables derived from family observable characteristics, while $\beta^s_{1}$ and $\boldsymbol{\beta}^s_{2}$, for $s\in\{u, ~m, ~v\}$, are the unknown parameters to be estimated. 
We control for a wide range of explanatory variables that may influence the network, such as parents' age, income, occupation, education, etc. (see household characteristics in Table \ref{tab:data}). As $m_{r,ij}$ and $v_{r,ik}$ are symmetric (Assumption \ref{ass:preferences}), we have $\boldsymbol{x}^{m}_{ij} = \boldsymbol{x}^{m}_{ji}$ and $\boldsymbol{x}^{v}_{ik} = \boldsymbol{x}^{v}_{ki}$. However, $\boldsymbol{x}^{u}_{ij}$ may not be symmetric. In other words, $\boldsymbol{x}^{u}_{ij}$ can encompass variables that drive homophily in linking decisions, as well as personal characteristics of families $i$ and $j$. In contrast, personal characteristics cannot be included in $\boldsymbol{x}^{m}_{ij}$ and $\boldsymbol{x}^{v}_{ik}$. For example, $\boldsymbol{x}^{u}_{ij}$ may include family $i$'s income, family $j$'s income, and the absolute difference in income, whereas $\boldsymbol{x}^{m}_{ij}$ and $\boldsymbol{x}^{v}_{ik}$ can only include the absolute difference in income. Additionally, as in Equation \eqref{eq:log}, we can also decompose the gender effect in \eqref{eq:spec} into effects specific to boys and girls.

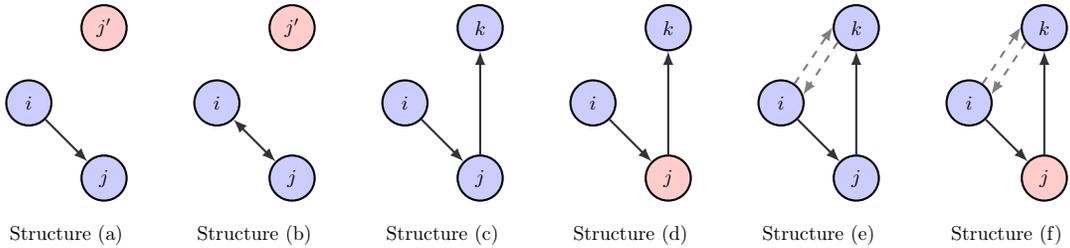
\begin{figure}[!htbp]
    \centering
    \begin{tikzpicture}[thick,scale=0.5, every node/.style={scale=0.75}]
            \tikzstyle{nodem}=[circle,draw,fill=blue!20,text=black, thick,minimum height=0.8cm]
            \tikzstyle{nodef}=[circle,draw,fill=red!20,text=black, thick,minimum height=0.8cm]
            \tikzstyle{linkr}=[->,>=latex, color=black!80, thick]
            \tikzstyle{linkb}=[<->,>=latex, color=black!80, thick]
            \tikzstyle{linkr2}=[->,>=latex, color=gray, thick, dashed]

            \node[nodem] (i1) at (0,0) {$i$};
            \node[nodem] (j1) at (2,-2) {$j$};
            \node[nodef] (jp1) at (2,2) {$j^{\prime}$};
            
            \draw[linkr] (i1)--(j1);
            \node[] at (1, -3.5){Structure (a)};

            \node[nodem] (i2) at (5,0) {$i$};
            \node[nodem] (j2) at (7,-2) {$j$};
            \node[nodef] (jp2) at (7,2) {$j^{\prime}$};

            \draw[linkb] (i2)--(j2);
            \node[] at (6, -3.5){Structure (b)};

            \node[nodem] (i3) at (10,0) {$i$};
            \node[nodem] (j3) at (12,-2) {$j$};
            \node[nodem] (k3) at (12,2) {$k$};

            \draw[linkr] (i3)--(j3);
            \draw[linkr] (j3)--(k3);
            \node[] at (11, -3.5){Structure (c)};

            \node[nodem] (i4) at (15,0) {$i$};
            \node[nodef] (j4) at (17,-2) {$j$};
            \node[nodem] (k4) at (17,2) {$k$};

            \draw[linkr] (i4)--(j4);
            \draw[linkr] (j4)--(k4);
            \node[] at (16, -3.5){Structure (d)};

            \node[nodem] (i5) at (20,0) {$i$};
            \node[nodem] (j5) at (22,-2) {$j$};
            \node[nodem] (k5) at (22,2) {$k$};

            \draw[linkr] (i5)--(j5);
            \draw[linkr] (j5)--(k5);
            \draw[linkr2] (i5.55)--(k5.178);
            \draw[linkr2] (k5.218)--(i5.20);
            \node[] at (21, -3.5){Structure (e)};

            \node[nodem] (i6) at (25,0) {$i$};
            \node[nodef] (j6) at (27,-2) {$j$};
            \node[nodem] (k6) at (27,2) {$k$};

            \draw[linkr] (i6)--(j6);
            \draw[linkr] (j6)--(k6);
            \draw[linkr2] (i6.55)--(k6.178);
            \draw[linkr2] (k6.218)--(i6.20);
            \node[] at (26, -3.5){Structure (f)};
        \end{tikzpicture}
    \caption{Illustration of the specification}
    \label{fig:spec}
\end{figure}

The influence of same-gender children is captured by the parameters $\beta_{1}^{u}$, $\beta_{1}^{m}$, and $\beta_{1}^{v}$. Figure \ref{fig:spec} presents potential situations that may arise in the network based on the signs of these parameters. In this visual representation, blue nodes represent families with boys, while red nodes represent families with girls. We are likely to observe structure (a) if $\beta_{1}^{u} > 0$, meaning that the network would contain more links (not necessarily reciprocal) between families with same-gender children. Conversely, if $\beta_{1}^{m} > 0$, structure (b) becomes more prevalent. In this scenario, families with children of the same gender are more inclined to form mutual connections. Finally, if $\beta_{1}^{v} > 0$, structures (c)--(f) are more probable. This suggests that families $i$ and $k$ are indirectly linked because they have boys, with this connection passing through family $j$, irrespective of the gender of family $j$ children, because $v(\boldsymbol{x}_{r,i}, \boldsymbol{x}_{r,k})$ is independent of $\boldsymbol{x}_{r,j}$.\footnote{As pointed out by \cite{mele2017structural}, it is possible to include $\boldsymbol{x}_{r,j}$ in $v(\boldsymbol{x}_{r,i}, \boldsymbol{x}_{r,k})$. However, $v(\boldsymbol{x}_{r,i}, \boldsymbol{x}_{r,j}, \boldsymbol{x}_{r,k})$ must be invariant to any permutation  of $\boldsymbol{x}_{r,i}$, $\boldsymbol{x}_{r,j}$, and $\boldsymbol{x}_{r,k}$ for the game to have a potential function.} In the model, the definition of indirect links from $i$ to $k$ does not preclude the existence of a direct link. This is the case, for example, in structures (e) and (f), where families $i$ and $k$ are connected both directly and indirectly in the network.

While the roles of the parameters $\beta_{1}^{u}$, $\beta_{1}^{m}$, and $\beta_{1}^{v}$ in the network are well-defined, their interpretation can be challenging. The complexity arises because the influence of same-gender children on direct links is not solely captured by $\beta_{1}^{u}$; it is also influenced by $\beta_{1}^{m}$ and $\beta_{1}^{v}$. For instance, when $\beta_{1}^{u} = 0$, it is not straightforward to conclude that same-gender children do not foster the formation of direct links. Even with $\beta_{1}^{u} = 0$, a positive value for $\beta_{1}^{m}$ corresponds to structure (b) in Figure \ref{fig:spec}, indicating a positive correlation between same-gender children and the likelihood to form a relationship. This is because a direct link in the network can also be involved in a mutual relationship. This duality is evident from Equation \eqref{eq:utility}, where $a_{r,ij}$ can take the value of 1 in both the direct and mutual utility terms for the same $i$ and $j$.  

Similarly, if $\beta_{1}^{u} = 0$ and $\beta_{1}^{m} = 0$, but $\beta_{1}^{v} > 0$, an increase in the proportion of same-gender children may result in a higher proportion of direct and mutual links. As illustrated in structures (e) and (f), on the one hand, family $k$ can act as both a direct (or mutual) and an indirect connection to $i$. Therefore, even if $\beta_1^u = 0$ and $\beta_1^m = 0$, we might still observe an increased probability of $i$ being directly connected to $k$ when $\beta_1^v > 0$. On the other hand, the indirect connection between $i$ and $k$ can go through a family, $j$, with a same-gender child, as is the case in structure (c). Increasing the prevalence of this structure also raises both the proportion of direct links and that of indirect links between families with same-gender children. The only scenario where an indirect link between families with same-gender children would not increase the proportion of direct links between these families is illustrated in structure (c). The actual impact of the condition $\beta_1^v > 0$ on the proportion of direct links between families with same-gender children depends on whether our dataset contains more instances of structures (c), (e), and (f) or structures (d).

The preceding discussion implies that observing $\beta_{1}^{u} = 0$ and $\beta_{1}^{m} = 0$ does not necessarily imply that same-gender children do not influence the likelihood of their families forming direct or mutual relationships. For this reason, we place less emphasis on interpreting the individual parameters in Equation \eqref{eq:spec}. Instead, our main interest lies in understanding the global effect of children's gender on the network. To achieve this, we conduct counterfactual analyses, varying, for instance, the proportion of boys in the sample. For each proportion, we measure specific network characteristics and analyze how they are influenced by children's gender.\footnote{One condition that would facilitate parameter interpretation is to define an indirect friend in the model as a friend of a friend who is not a friend and to refer to a direct friend as a non-mutual friend. Under these definitions, there should be no overlap between structure (c) and structures (a) and (b), nor should structure (b) overlap with structure (a). However, by imposing these restrictions, the game would have no potential function, making the identification of parameters impossible.}


\paragraph{Parameter identification}
In our dataset, 65.43\% of direct links between families with same-gender children are involved in mutual links. Additionally, 51.73\% of indirect links between families with same-gender children pass through a family with a same-gender child, and 66.45\% of these indirect links are also direct.\footnote{These proportions indicate that indirect links between families with same-gender children generate a substantial number of direct links between these families. On one hand, there are 116,668 indirect links between families with same-gender children (note that we may observe more than one indirect link between two families as the link can go through many families). Given this large number of indirect links, the proportion of 51.73\% is statistically greater than 50\%, suggesting that indirect links between families with same-gender children often go through a family with a same-gender child. On the other hand, the proportion of 66.45\% indicates that these indirect links are also direct links in most cases.} This observation reinforces our earlier point that $\beta_{1}^{v}$ captures the influence of same-gender children on the probability of forming both direct and mutual links. Importantly, this situation may also give rise to a non-identification issue. Specifically, we are concerned that many combinations of $\beta_{1}^{u}$, $\beta_{1}^{m}$, and $\beta_{1}^{v}$ could produce the same network structure. If this is the case, it might be challenging to disentangle between the parameters.

The non-identification issue in this model has also been highlighted by \cite{mele2017structural}. For instance, in Figure \ref{fig:net}, we can typically replicate structures (c)--(f) using only direct and mutual links between multiple nodes. This implies that, in certain cases, we can achieve the same network configuration by increasing $\beta_{1}^{u}$ and $\beta_{1}^{m}$ while decreasing $\beta_{1}^{v}$. Links that will break up in some indirect relationships due to the reduction in $\beta_{1}^{v}$ could be regenerated as direct and mutual links due to the increase in $\beta_{1}^{u}$ and $\beta_{1}^{m}$.

Consider a scenario where we observe a single village in which all families are mutually connected. In this particular situation, it becomes impossible to distinguish between the parameters $\beta_{1}^{u}$, $\beta_{1}^{m}$, and $\beta_{1}^{v}$. However, if we observe a second village where no links are mutual, it would not be possible to achieve the same network structures using various combinations of $\beta_{1}^{u}$ and $\beta_{1}^{m}$. Therefore, a sufficient condition for identifying the individual effects of $\beta_{1}^{u}$ and $\beta_{1}^{m}$ is the presence of enough variation in the proportion of direct links between families with same-gender children that are also mutual across villages. In addition, to identify $\beta_{1}^{v}$ from the rest of the parameters, we also need the proportion of direct and mutual links between families having same-gender children involved by indirect relation between these families to vary sufficiently across villages. Fortunately, this condition is likely to be met because we have a substantial dataset with 222 villages assumed to be independent. Table \ref{tab:propolink} provides an overview of the distribution of link proportions among families with same-gender children across different villages. The significant variability in these proportions ensures the identification of model parameters. This identification analysis can also be extended to parameters associated with other control variables in Equation \eqref{eq:spec}. Using variation across groups to achieve identification is also employed by \cite{mele2022structural} in a context where networks from multiple schools are assumed to be independent.

\begin{table}[!ht]
\footnotesize
\centering
\caption{Link proportions between families with same-gender children across villages}
\begin{threeparttable}
\begin{tabular}{ld{4}d{4}d{4}d{4}d{4}}
\toprule
                                                                   & \multicolumn{1}{c}{Min}   & \multicolumn{1}{c}{Pctl 25} & \multicolumn{1}{c}{Median} & \multicolumn{1}{c}{Pctl 75} & \multicolumn{1}{c}{Max}   \\ \midrule
Direct links that are also mutual                                       & 0.133 & 0.522   & 0.644  & 0.735   & 1.000 \\
Indirect links through families with same-gender children & 0.000 & 0.418   & 0.483  & 0.590   & 1.000 \\
Indirect links that are also non-mutually direct                    & 0.000 & 0.161   & 0.211  & 0.259   & 0.643 \\
Indirect links that are also mutually direct                                      & 0.048 & 0.271   & 0.398  & 0.509   & 1.000 \\
Indirect links that involve direct reciprocal (non-mutual) links       & 0.000 & 0.077   & 0.109  & 0.135   & 0.301 \\ \bottomrule
\end{tabular}
\end{threeparttable}
\begin{tablenotes}[para,flushleft]
\footnotesize
This table summarizes link proportions between families with same-gender children across villages. Pctl 25 and Pctl 75 represent the 25th and 75th percentiles of these proportions.  ``Indirect links that involve direct reciprocal (non-mutual) links" means that $i$ is indirectly connected to $k$ with a direct (non-mutual) link from $k$ to $i$.
\end{tablenotes}
\label{tab:propolink}
\end{table}

\paragraph{Estimation strategy}

Our estimation strategy follows \cite{mele2022structural}. Specifically, we adapt the Bayesian estimation approach proposed by \cite{mele2017structural} to our context where we have a large number of villages. This approach relies on a Markov Chain Monte Carlo (MCMC) process to simulate the posterior distribution of the parameters. Importantly, this approach circumvents the need to compute the denominator in the distribution of $A_r$ as given by Equation \eqref{eq:pi}. This denominator involves summing over all possible networks, and, as we pointed out earlier, this can amount to millions of networks even with a relatively small number of families per village. 

In the MCMC process, a computationally intensive task is to simulate, in each iteration, an artificial network $\tilde A_r$ with the same likelihood as that of the observed network $A_r$. As the artificial network must be independent across villages as the observed network, we perform this simulation for multiple villages simultaneously. Our MCMC process consists of 100,000 iterations, with the last 50,000 iterations retained for estimating the posterior distribution of the parameters. During each iteration, we conduct 7,000 simulations of artificial networks and retain the final one, as suggested by \cite{mele2017structural}. To achieve an acceptance rate of 23.4\%, we propose an adaptive version, as detailed in Appendix \ref{append:est}, following the approach outlined in \cite{atchade2005adaptive}.

For the counterfactual analysis, we perform 1,000 simulations for each assigned value of the network explanatory variables.\footnote{To assign a value to the explanatory variables, one approach is to generate, for example, the variable ``Child Gender'' to achieve a specific proportion of same-gender children in the sample, while keeping other control variables fixed.} In each simulation, we draw parameter values from the last 50,000 simulations of the MCMC. Once these values are drawn, given the assigned values of the explanatory variables, we perform 7,000 network simulations in each village. We retain the last simulations to compute network characteristics, such as degree, clustering density, diameter, etc. Therefore, given the value assigned to the network explanatory variables, we have 1,000 simulations for each network characteristic that can be used to estimate the average and confidence interval of the characteristic (see Appendix \ref{append:countf}).

\section{Results\label{sec:result}}

\subsection{Posterior estimates}

Table \ref{TAB:Post1}  presents the estimated posterior means that we obtain from structural estimation of the preference parameters of the model. This table contains two specifications: the first specification only considers direct links, while the second specification includes mutual and indirect links as well.\footnote{For completeness, we present two additional specifications in Appendix Table \ref{TAB:Post2}: one that only includes benefits from mutual links and another that includes both direct and mutual links (but not indirect ones).} Following the presentation in Table \ref{TAB:logit}, in columns (1) and (3), we model the effect of a pair of families having same-gender children, while in columns (2) and (4), we allow having boys to have different effect than having girls. As in our previous descriptive analysis, the number of observations in these estimations corresponds to the 92,152 possible pairs of links that the 4,473 families in the sample can form. Each estimated coefficient that we report in Table \ref{TAB:Post1} represents the marginal impact of the associated variable on the utility of forming a direct, mutual, or indirect link, respectively, while other variables are kept constant. Our primary focus in discussing these results revolves around the effect of having same-gender children ($Gender_{ij}$), which we then decompose to the effect of both families having boys and the effect of both families having girls. All specifications also include various demographic controls (parents' age, income, occupation, and education) for families $i$ and $j$.

Starting with the estimates reported in columns (1) and (2), we observe that the sign and statistical significance of the coefficients are very similar to those reported using a standard logit model of link formation in Table \ref{TAB:logit}, although the estimation approach is different. This provides some confidence in the model and Bayesian estimation approach pursued here.

Our preferred specification, of course, is one which also accounts for preferences for mutual and indirect links presented in columns (3) and (4). This specification also captures equilibrium and network effects of network formation that a standard logit model cannot address. In column (3), we observe that the coefficients on ``$\textit{Child Gender}_{ij}$'' are positive across all three terms of the specification, although only the coefficient on indirect links is statistically significant. Note, however, that indirect links encompass the benefits of both direct and mutual links, as these are prerequisites for indirect links to be formed. In addition, direct links in the model can be part of mutual relationships. As pointed out earlier, we observe that 65.43\% of direct links between families with same-gender children are involved in mutual links. Additionally, 51.73\% of indirect links between families with same-gender children pass through a family with a same-gender child and 66.45\% of these indirect links are also direct. The significance of same-gender children on indirect link formation indicates that not only does the gender of children of families that are one-step away connections matter for network formation but more widely of the community of families that are connected to these direct links. It is also worth highlighting that the indirect links we consider here can also be direct links, meaning that $i$ can be connected to $j$ and $k$, and $j$ also connected to $k$, so from $i$'s perspective $j$ and $k$ can be considered as both direct and indirect links (same applies to $j$ and $k$). This makes the interpretation of the coefficients we estimate complicated, which is why turning to counterfactual simulations is insightful, as it allows us to trace out the full implications of changes in one of the factors on the structure of networks.

In terms of other characteristics, several noteworthy observations stand out: income gaps between families play a significant role for both direct and mutual links. Age gaps between fathers and age gaps between mothers seem to have significant and opposing effects (for mutual and indirect links). Having similar education levels is important for indirect links and having similar occupation has ambiguous effects.

\renewcommand{\arraystretch}{0.9}
\begin{table}[htbp]
  \centering 
  \footnotesize
  \addtolength{\leftskip} {-0.5cm}
  \caption{Summary of the parameter posterior distribution} 
\resizebox{1.1\textwidth}{!}{
\begin{threeparttable}
\begin{tabular}{ld{7}ld{7}ld{7}ld{7}l}
\toprule
                                      & \multicolumn{2}{c}{(1)} & \multicolumn{2}{c}{(2)} & \multicolumn{2}{c}{(3)} & \multicolumn{2}{c}{(4)} \\ \midrule
\textbf{Direct links}                 & \textbf{}      & \textbf{}  & \textbf{}      & \textbf{}  & \textbf{}      & \textbf{}  & \textbf{}      & \textbf{}  \\
Intercept                             & -1.003^{***} & (0.069)    & -1.005^{***} & (0.067)    & -2.257^{***}        & (0.070)    & -2.261^{***}       & (0.073)    \\
$\text{Child Gender}_{ij}$                  & 0.021^{}     & (0.015)    &                                 &            & 0.010^{}            & (0.025)    &                                       &            \\
$\text{Child Male}_{ij}$                    &                                 &            & 0.043^{**}   & (0.018)    &                                        &            & 0.020^{}           & (0.026)    \\
$\text{Child Female}_{ij}$                  &                                 &            & 0.012^{}     & (0.017)    &                                        &            & -0.026^{}          & (0.026)    \\
Adiff(Father’s Age)                   & 0.002^{}     & (0.002)    & 0.002^{}     & (0.002)    & \textless{}0.000^{} & (0.003)    & \textless 0.000^{} & (0.003)    \\
$\text{Age Father}_i$                 & -0.002^{}    & (0.002)    & -0.002^{}    & (0.002)    & 0.001^{}            & (0.002)    & -0.002^{}          & (0.002)    \\
$\text{Age Father}_j$                 & -0.005^{**}  & (0.002)    & -0.005^{**}  & (0.002)    & -0.005^{**}         & (0.002)    & -0.003^{*}         & (0.002)    \\
Adiff(Mother’s Age)                   & -0.005^{**}  & (0.002)    & -0.005^{**}  & (0.002)    & -0.007^{**}         & (0.003)    & -0.006^{*}         & (0.003)    \\
$\text{Age Mother}_i$                 & 0.002^{}     & (0.002)    & 0.002^{}     & (0.003)    & -0.001^{}           & (0.003)    & 0.003^{}           & (0.002)    \\
$\text{Age Mother}_j$                 & 0.006^{**}   & (0.003)    & 0.006^{**}   & (0.002)    & 0.007^{***}         & (0.003)    & 0.004^{***}        & (0.002)    \\
Adiff(Income)                         & -0.016^{***} & (0.002)    & -0.016^{***} & (0.002)    & -0.008^{***}        & (0.002)    & -0.008^{***}       & (0.002)    \\
$\text{Income}_i$                     & 0.012^{***}  & (0.002)    & 0.012^{***}  & (0.002)    & 0.008^{***}         & (0.002)    & 0.008^{***}        & (0.002)    \\
$\text{Income}_j$                     & 0.013^{***}  & (0.002)    & 0.013^{***}  & (0.002)    & 0.010^{***}         & (0.002)    & 0.010^{***}        & (0.002)    \\
Adiff(Homestead land)                 & -0.408^{***} & (0.121)    & -0.418^{***} & (0.125)    & 0.060^{}            & (0.121)    & 0.072^{}           & (0.113)    \\
$\text{Homestead land}_i$             & 0.304^{***}  & (0.116)    & 0.305^{***}  & (0.117)    & 0.492^{***}         & (0.109)    & 0.483^{***}        & (0.097)    \\
$\text{Homestead land}_j$             & 0.187^{*}    & (0.111)    & 0.203^{*}    & (0.116)    & 0.307^{***}         & (0.106)    & 0.299^{***}        & (0.111)    \\
Adiff(Farming land)                   & -0.185^{***} & (0.018)    & -0.187^{***} & (0.017)    & -0.112^{***}        & (0.015)    & -0.110^{***}       & (0.014)    \\
$\text{Farming land}_i$               & 0.179^{***}  & (0.017)    & 0.181^{***}  & (0.017)    & 0.110^{***}         & (0.015)    & 0.107^{***}        & (0.013)    \\
$\text{Farming land}_j$               & 0.190^{***}  & (0.017)    & 0.191^{***}  & (0.017)    & 0.125^{***}         & (0.015)    & 0.122^{***}        & (0.013)    \\
$\text{Father’s education}_{ij}$      & 0.055^{***}  & (0.016)    & 0.055^{***}  & (0.015)    & 0.010^{}            & (0.028)    & 0.003^{}           & (0.023)    \\
$\text{Mother’s education}_{ij}$      & 0.031^{**}   & (0.014)    & 0.031^{**}   & (0.015)    & -0.021^{}           & (0.024)    & -0.008^{}          & (0.027)    \\
$\text{Father is daily laborer}_{ij}$ & 0.268^{***}  & (0.021)    & 0.267^{***}  & (0.021)    & 0.021^{}            & (0.038)    & 0.016^{}           & (0.041)    \\
$\text{Father is farmer}_{ij}$        & 0.151^{***}  & (0.030)    & 0.152^{***}  & (0.030)    & 0.026^{}            & (0.053)    & 0.033^{}           & (0.043)    \\
$\text{Father is self-employed}_{ij}$ & -0.099^{***} & (0.024)    & -0.096^{***} & (0.024)    & 0.079^{**}          & (0.040)    & 0.088^{**}         & (0.039)    \\
$\text{Mother stays home}_{ij}$       & -0.073^{***} & (0.020)    & -0.072^{***} & (0.021)    & -0.080^{**}         & (0.037)    & -0.067^{*}         & (0.038)    \\\midrule
\textbf{Mutual links}                 &                                 &            &                                 &            &                                        &            &                                       &            \\
Intercept                             &                                 &            &                                 &            & 2.093^{***}         & (0.074)    & 2.124^{***}        & (0.060)    \\
$\text{Gender}_{ij}$                  &                                 &            &                                 &            & 0.002^{}            & (0.050)    &                                       &            \\
$\text{Male}_{ij}$                    &                                 &            &                                 &            &                                        &            & 0.008^{}           & (0.048)    \\
$\text{Female}_{ij}$                  &                                 &            &                                 &            &                                        &            & -0.005^{}          & (0.047)    \\
Adiff(Father’s Age)                   &                                 &            &                                 &            & -0.006^{}           & (0.005)    & -0.007^{}          & (0.005)    \\
Adiff(Mother’s Age)                   &                                 &            &                                 &            & 0.014^{**}          & (0.007)    & 0.013^{**}         & (0.007)    \\
Adiff(Income)                         &                                 &            &                                 &            & -0.011^{***}        & (0.004)    & -0.010^{**}        & (0.004)    \\
Adiff(Homestead land)                 &                                 &            &                                 &            & -0.586^{***}        & (0.164)    & -0.606^{***}       & (0.179)    \\
Adiff(Farming land)                   &                                 &            &                                 &            & -0.015^{}           & (0.010)    & -0.014^{}          & (0.010)    \\
$\text{Father’s education}_{ij}$      &                                 &            &                                 &            & 0.034^{}            & (0.053)    & 0.045^{}           & (0.046)    \\
$\text{Mother’s education}_{ij}$      &                                 &            &                                 &            & 0.066^{}            & (0.048)    & 0.033^{}           & (0.047)    \\
$\text{Father is daily laborer}_{ij}$ &                                 &            &                                 &            & -0.108^{}           & (0.068)    & -0.104^{}          & (0.075)    \\
$\text{Father is farmer}_{ij}$        &                                 &            &                                 &            & 0.094^{}            & (0.099)    & 0.074^{}           & (0.078)    \\
$\text{Father is self-employed}_{ij}$ &                                 &            &                                 &            & 0.146^{**}          & (0.071)    & 0.142^{**}         & (0.065)    \\
$\text{Mother stays home}_{ij}$       &                                 &            &                                 &            & 0.060^{}            & (0.058)    & 0.042^{}           & (0.057)    \\
\\\midrule
\textbf{Indirect links}               &                                 &            &                                 &            &                                        &            &                                       &            \\
Intercept                             &                                 &            &                                 &            & 0.026^{***}         & (0.004)    & 0.025^{***}        & (0.003)    \\
$\text{Gender}_{ij}$                  &                                 &            &                                 &            & 0.007^{**}          & (0.003)    &                                       &            \\
$\text{Male}_{ik}$                    &                                 &            &                                 &            &                                        &            & 0.005^{**}         & (0.002)    \\
$\text{Female}_{ik}$                  &                                 &            &                                 &            &                                        &            & 0.011^{***}        & (0.002)    \\
Adiff(Father’s Age)                   &                                 &            &                                 &            & 0.001^{***}         & (0.000)    & 0.001^{***}        & (0.000)    \\
Adiff(Mother’s Age)                   &                                 &            &                                 &            & -0.001^{***}        & (0.000)    & -0.001^{***}       & (0.000)    \\
Adiff(Income)                         &                                 &            &                                 &            & 0.215^{}            & (0.190)    & 0.130^{}           & (0.192)    \\
Adiff(Homestead land)                 &                                 &            &                                 &            & -0.049^{***}        & (0.009)    & -0.047^{***}       & (0.008)    \\
Adiff(Farming land)                   &                                 &            &                                 &            & -0.073^{}           & (0.700)    & 0.001^{}           & (0.681)    \\
$\text{Father’s education}_{ik}$      &                                 &            &                                 &            & 0.006^{**}          & (0.003)    & 0.005^{**}         & (0.002)    \\
$\text{Mother’s education}_{ik}$      &                                 &            &                                 &            & 0.008^{***}         & (0.002)    & 0.008^{***}        & (0.002)    \\
$\text{Father is daily laborer}_{ik}$ &                                 &            &                                 &            & 0.031^{***}         & (0.003)    & 0.031^{***}        & (0.003)    \\
$\text{Father is farmer}_{ik}$        &                                 &            &                                 &            & 0.002^{}            & (0.003)    & 0.003^{}           & (0.003)    \\
$\text{Father is self-employed}_{ik}$ &                                 &            &                                 &            & -0.034^{***}        & (0.004)    & -0.036^{***}       & (0.004)    \\
$\text{Mother stays home}_{ik}$       &                                 &            &                                 &            & -0.001^{}           & (0.003)    & -0.001^{}          & (0.003)   \\\bottomrule
\end{tabular}
\begin{tablenotes}[para,flushleft]
\footnotesize
Note: ***: \textit{pvalue $<1\%$}, **: \textit{pvalue $<5\%$}, *: \textit{pvalue $<10\%$}\\
This table presents the mean of the posterior distribution of each parameter followed by the standard deviation in brackets using the last 50,000 simulations of the MCMC. For any variable denoted ``var", ``var$_{ij}$'' takes one if var$_i =$ var$_j$ and zero otherwise; ``Adiff(var)'' if the absolute difference between var$_i$ and var$_j$. The network is directed and the relationship we model is the link from $j$ to $i$.
\end{tablenotes}
\end{threeparttable}}
\label{TAB:Post1}
\end{table}
\renewcommand{\arraystretch}{1}

\subsection{Counterfactual experiments}

We next use the estimated model to perform counterfactual experiments that enable us to gauge the impact of changes in characteristics of interest on several key network structure measures: degree (number of links), asymmetry, density, average distance, and clustering. Asymmetry measures variance in the decision to form a link between two families chosen arbitrarily, that is, the binary variable $a_{r,ij}$. Higher values indicate disparity in the decision to form a relationship. We normalize the asymmetry for the highest value to be equal to one. Network density represents the proportion of actual connections within a network relative to all potential connections. Its values range between 0 and 1, with higher values indicating denser networks. Network clustering captures the tendency of families in a network to form clusters (triangles), and ranges from 0 (no clustering) to 1 (maximum clustering). To be precise, it measures the ratio of fully connected triples to the potential triples that include at least two links.

\begin{figure}[!htbp]
    \centering
    \begin{subfigure}{\textwidth} 
        \centering
        \includegraphics[scale=1, width=\textwidth]{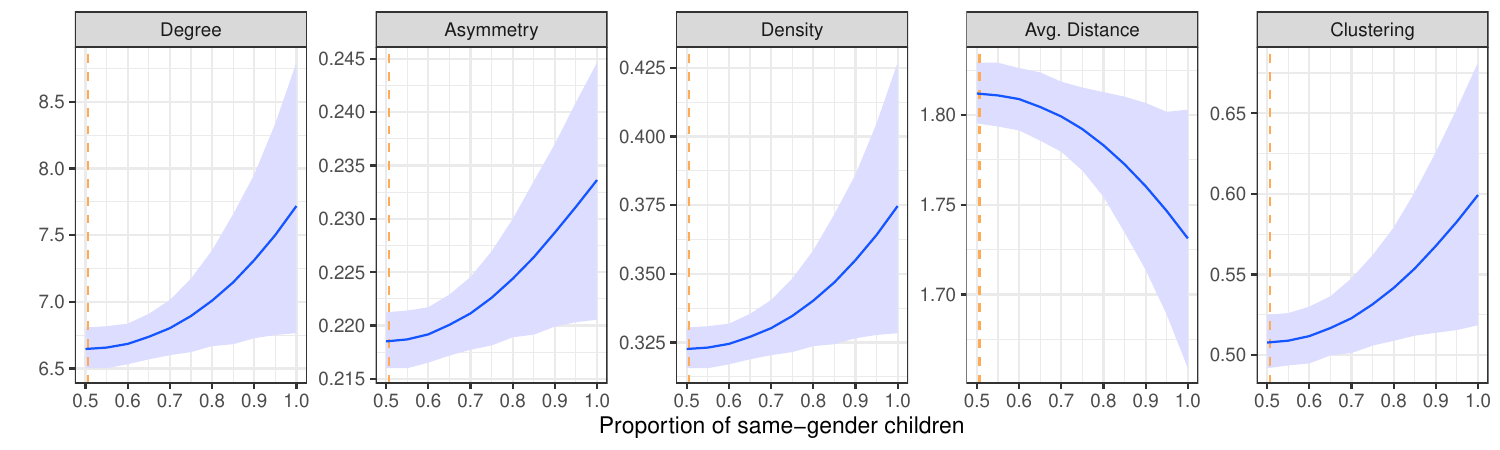}
        \caption{Impact of the proportion of same-gender children}
     
    \end{subfigure}
    
    \bigskip 
    
    \begin{subfigure}{\textwidth}
        \centering
        \includegraphics[scale=1, width=\textwidth]{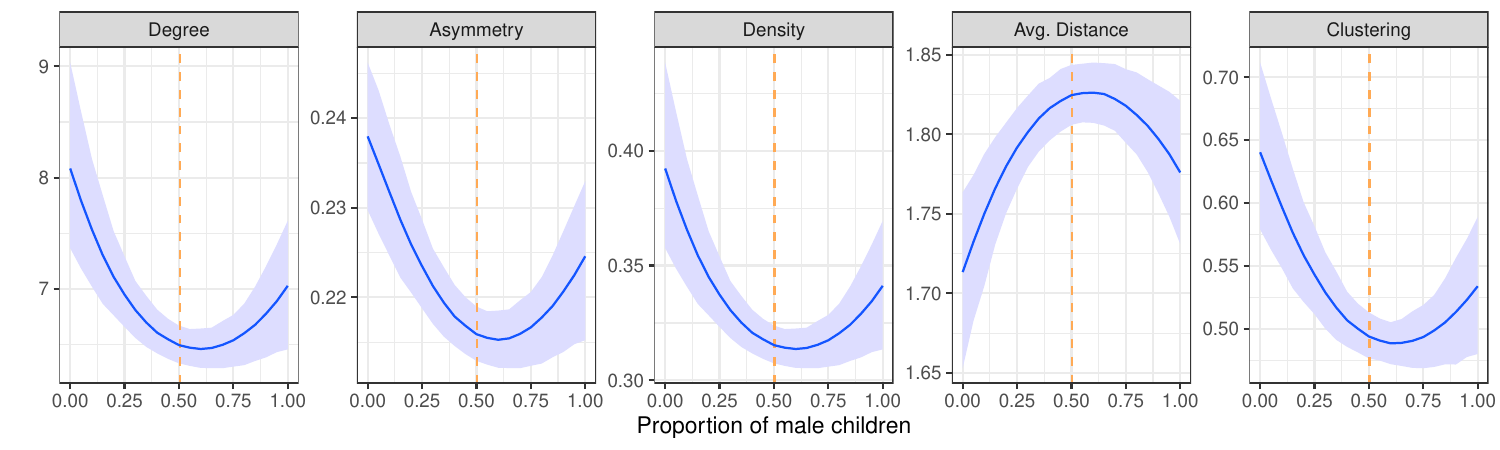}
        \caption{Impact of the proportion of male children}

    \end{subfigure}

    \medskip 
    
    \begin{minipage}{\textwidth}
    \footnotesize
    \textbf{Notes:}
    
    \textbf{Panel (a):} Children's gender is randomly generated as a dummy variable from a Bernoulli distribution. The proportion of same-gender children varies from 50\% (meaning that they are 50\% boys) to 100\% (meaning that all the children are either boys or girls).
    
    \textbf{Panel (b):} Child being male or not is randomly generated as a dummy variable from a Bernoulli distribution of parameter the proportion of male children desired in the sample. Changing the variable ``Male'' also requires changing the variable ``Female,'' which is defined as one minus the variable ``Male.''
    \end{minipage}
       \caption{Counterfactual experiment on gender composition of children.}
          \label{fig:simugender}
\end{figure}

Our counterfactual analysis is based on 1,000 independent draws from the set constituted by the last 50,000 simulations on the MCMC process. For each draw, we generate the network by varying the explanatory variable depending on the targeted analysis. For example, in Figure \ref{fig:simugender}, we vary the proportion of same-gender children from 0\% to 100\%. We then plot various network characteristics, including degree, asymmetry, density, average distance, and clustering.

\paragraph{Children's Gender}

Figure \ref{fig:simugender} illustrates our first counterfactual experiment in which we vary the gender composition of the children in the sample. In panel (a), we vary the gender composition of children from being split to being fully homogeneous (either boys or girls), whereas in panel (b), we vary the proportion of male children from 0 to 1 to assess whether there are gender differences in the impact.

Starting with panel (a), we observe that as the gender composition becomes more homogeneous, there is a significant impact on all the measures of network structure that we consider. Notably, the average number of links (i.e., degree) increases by about 15\%  and, therefore, the network density and asymmetry go up as well, along with a reduction in the average distance between two nodes. We also see a significant increase in clustering by about 20\%  and a modest increase in asymmetry.

When we turn our attention to panel (b), we observe that girls have a stronger effect on network structure. Comparing the scenario of all girls to either the status quo or the scenario of all boys, we observe that the average degree is higher by about 21\% in the first case and 14\% in the second case. Similarly, girls have a stronger impact on the other dimensions, suggesting that they give rise to denser and more clustered networks than the status quo or the all-boys scenario.

These results highlight a significant role of children's gender for parents' networks; however, they do not provide a precise sense of the effect's magnitude. To establish a benchmark against which we can gauge the quantitative importance of children's gender, we next present counterfactual experiments in which we vary other important determinants of network formation, such as households' income, fathers' occupation, fathers' and mothers' education and age. 

\paragraph{Parents' Income}

\begin{figure}[!htbp]
    \centering
    \includegraphics[scale = 1, width=\textwidth]{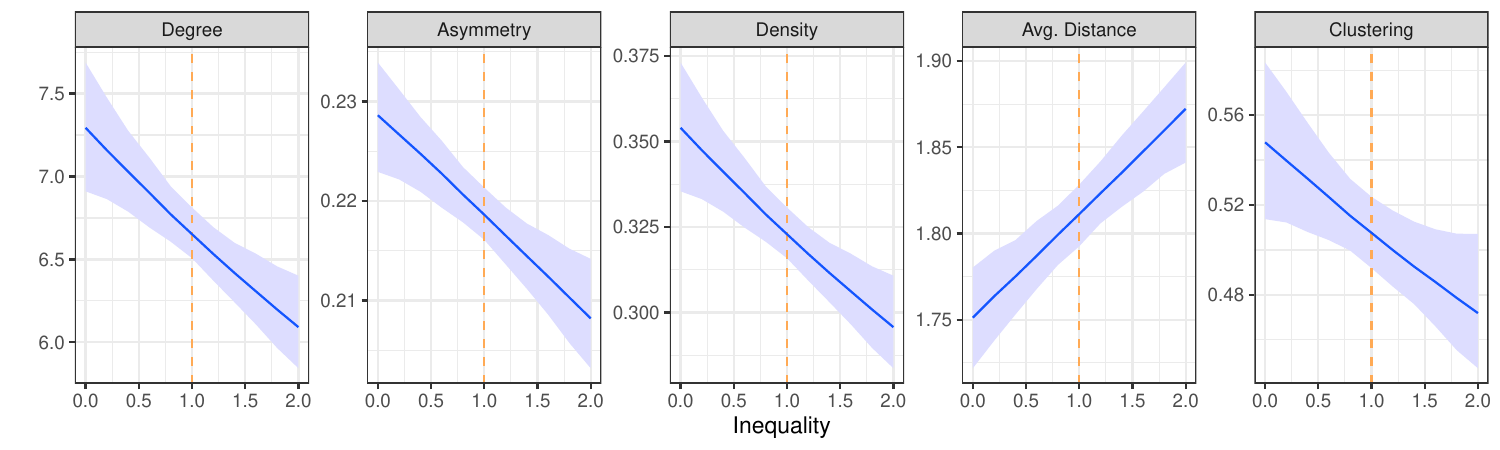}
    \caption{Impact of income inequality}
    \label{fig:simuinc}

    \justify{\footnotesize Note: For each simulation, we vary the income inequality by generating a new income variable given by $\hat{\text{Inc}} = \text{Inc}_0 + \Bar{\delta}\times (\text{Inc} - \text{Inc}_0)$, where $\text{Inc}$ in the observed income in the original sample, $\text{Inc}_0$ is the average income observed in the original sample, and $\Bar{\delta} $ is the inequality measure. If $\Bar{\delta} = 0$, then the generated income is the observed average income, i.e. the generated income is the same for all families. If $\Bar{\delta} = 1$, then the generated income coincides with the observed income. If $\Bar{\delta} = 2$, then the standard deviation of the generated income is twice that of the observed income.}
\end{figure}

Figure \ref{fig:simuinc} presents our second counterfactual experiment in which we vary income inequality across households. Specifically, the horizontal access measures income inequality, with a 0 corresponding to the scenario of perfect equality, 1  to the observed income, and 2 to a scenario of high inequality (twice that of the observed one). All measures vary monotonically with inequality in the direction of families having fewer connections, and networks becoming less dense and clustered. It is of interest to consider what are the quantitative implications of income inequality and how do they compare to those of children's gender. 
By moving from the status quo (1) to perfect equality (0), we would increase the average degree by about 7.3, which is within the range\textemdash in fact, slightly less\textemdash than the all-girls scenario depicted  in Figure \ref{fig:simugender}. Thus, our first take-away is that children's gender is as important for the structure of parental networks as income distribution.

\paragraph{Fathers' Occupation}

\begin{figure}[!htbp]
    \centering
    \begin{subfigure}{\textwidth} 
        \centering
    \includegraphics[scale = 1, width=\textwidth]{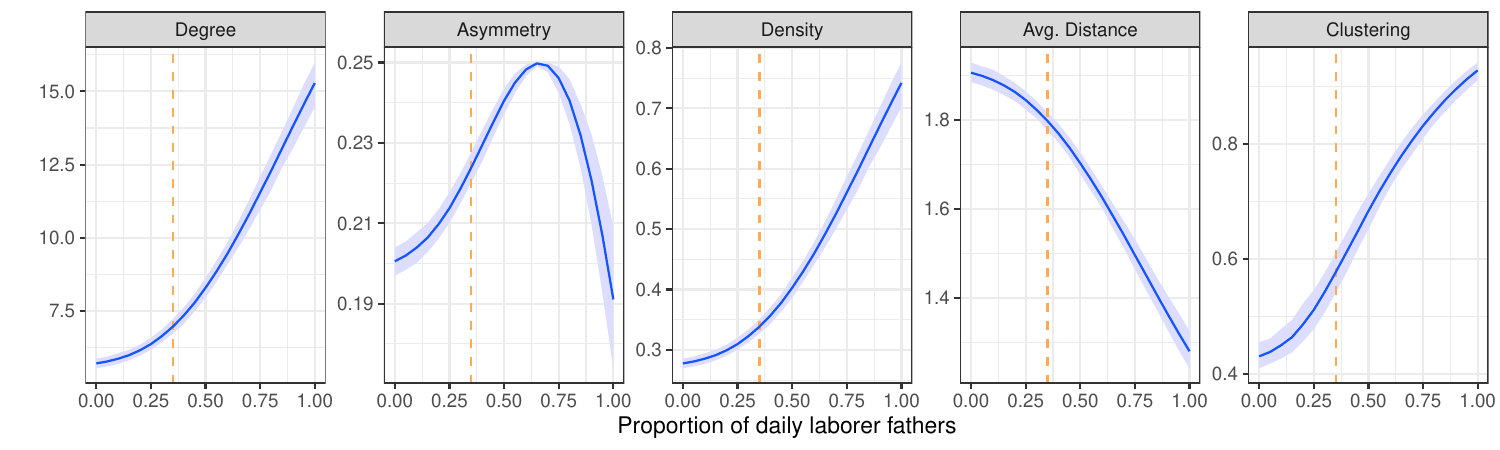}
    \caption{Impact of the proportion of daily laborer fathers}
     
    \end{subfigure}
    
    \bigskip 
    
    \begin{subfigure}{\textwidth}
        \centering
     \includegraphics[scale = 1, width=\textwidth]{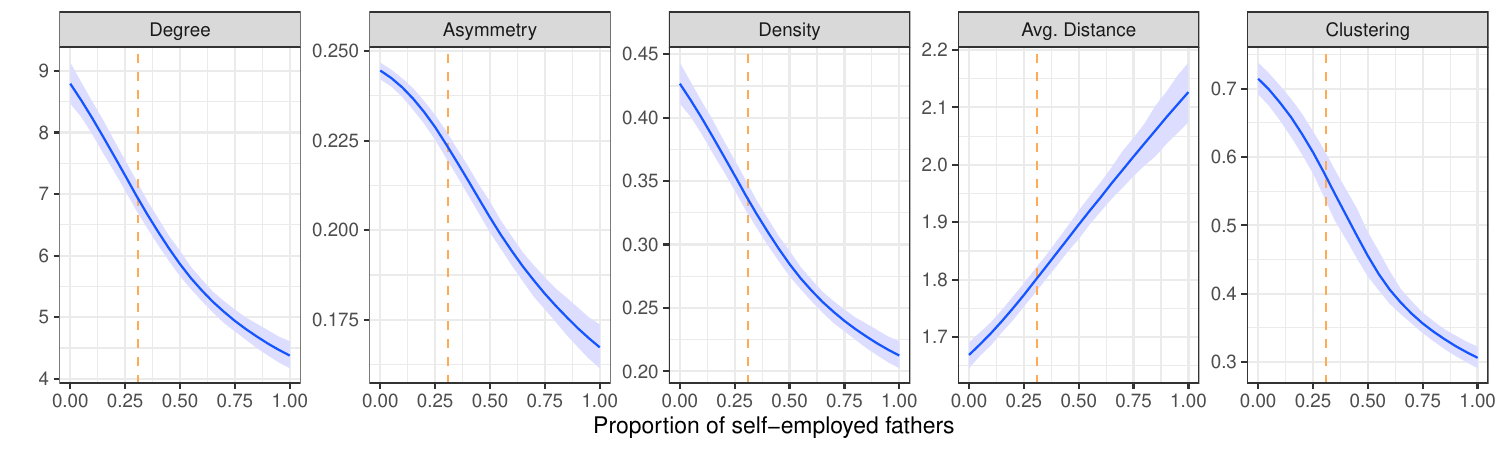}
    \caption{Impact of the proportion of self-employed fathers}
       
    \end{subfigure}
       
    \medskip 
    
    \begin{minipage}{\textwidth}
    \footnotesize
    \textbf{Notes:} Father being a daily laborer (self-employed) is generated as a dummy variable from a Bernoulli distribution. Changing the variable "daily laborer" ("self-employed") also requires changing the proportion of other occupations. Once the variable "daily laborer fathers" ("self-employed") is generated, we proportionally adjust the proportion of the other occupations for the total proportion to equal one.
    \end{minipage}
       \caption{Counterfactual experiment on occupation of fathers.}
          \label{fig:simuoccupation}
\end{figure}

Our third counterfactual experiment focuses on varying the heterogeneity of    fathers' occupations across families. We select two occupations that are most prevalent in the sample\textemdash that is, daily laborers and self-employed\textemdash and we vary the proportion of fathers who have these   occupations. Panel (a) in Figure \ref{fig:simuoccupation} displays the case of daily laborers.  It is evident that more homogeneity has a strong effect on network degree, since moving from the status quo (marked by the vertical dashed line) to full homogeneity\textemdash that is, all fathers are daily laborers\textemdash  leads to a doubling of the average degree, the network density, and the overall clustering. In contrast, asymmetry exhibits different behavior due to the significant increase in degree.\footnote{Asymmetry measures the variance of the binary variable $a_{r,ij}$, which represents the decision to form a relationship. The variance is highest when the probability of $a_{r,ij}$ being one reaches 0.5. When this probability exceeds 0.5, the variance decreases. This is in line with the asymmetry behavior. When the degree is below 10, which is half of the average village size, asymmetry increases with the degree. Conversely, when the degree exceeds half of the average village size, asymmetry decreases.} 

On the other hand, the picture is quite different in the case of self-employment. As we increase the proportion of self-employed fathers, we observe in panel (b) of Figure \ref{fig:simuoccupation} a significant and monotonic reduction in degree, density, and clustering. This may be because self-employed individuals in this village context may be small shop-keepers or engaged in handicraft production. These occupations might place them in competition with each other in the village market, suggesting that they might be less likely to form strong connections with each other for risk-sharing and mutual support purposes.

In Figure \ref{fig:simufarmers} in the Appendix, we also present a similar experiment on varying proportion of farmers, which has the same qualitative features as that of daily laborers, though the magnitudes are lower.

\paragraph{Parents' Education}

\begin{figure}[!htbp]
    \centering
    \begin{subfigure}{\textwidth} 
        \centering
    \includegraphics[scale = 1, width=\textwidth]{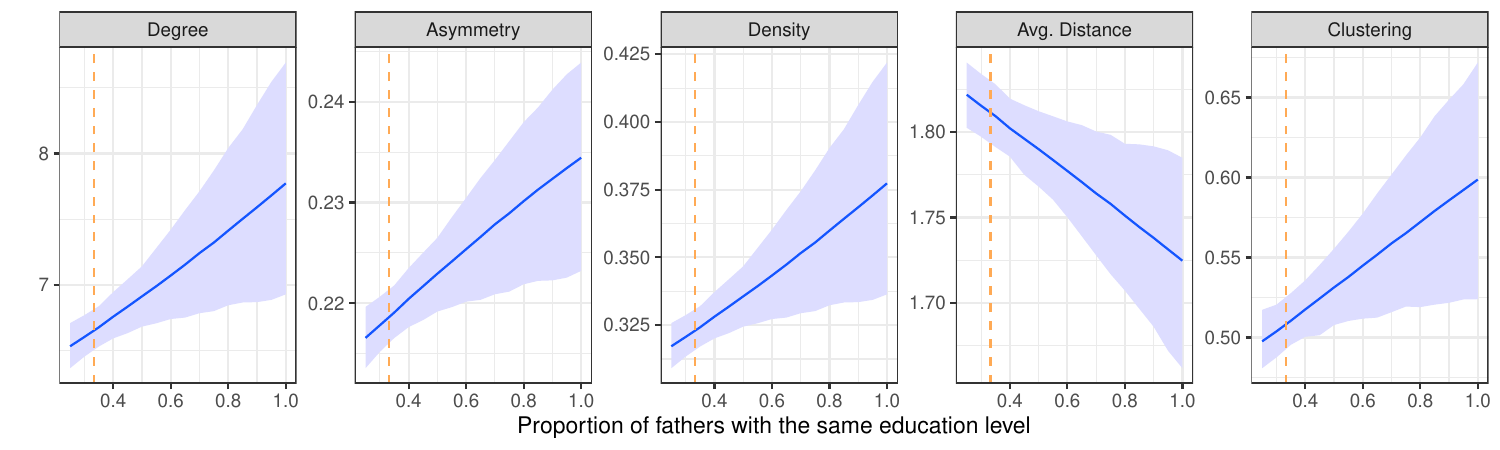}
    \caption{Impact of the proportion of fathers with the same education level}
     
    \end{subfigure}
    
    \bigskip 
    
    \begin{subfigure}{\textwidth}
        \centering
    \includegraphics[scale = 1, width=\textwidth]{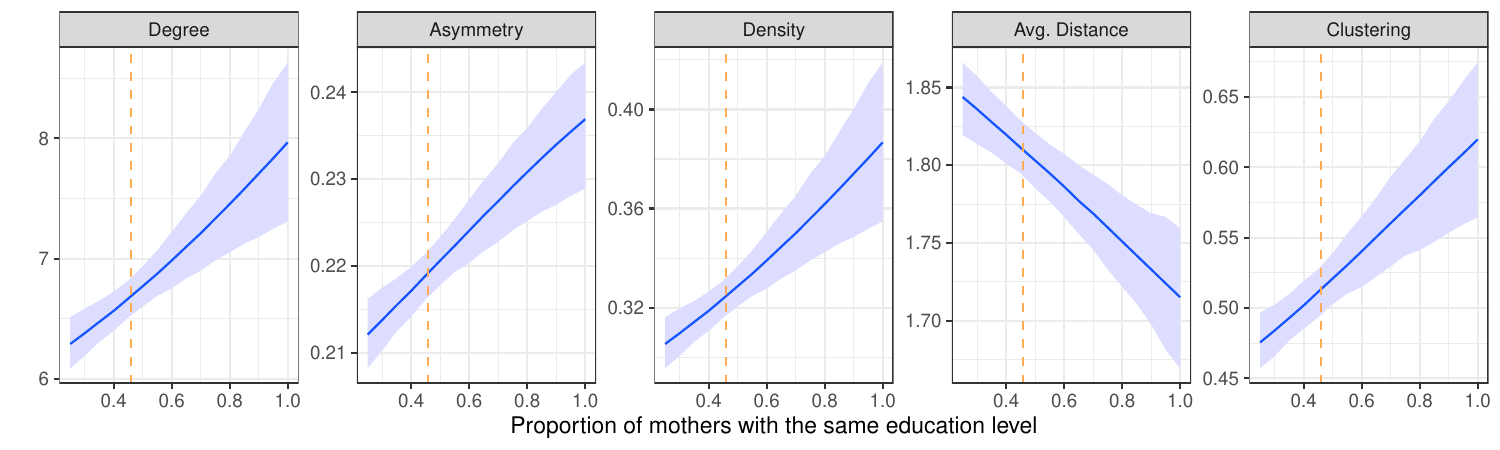}
    \caption{Impact of the proportion of mothers with the same education level}
       
    \end{subfigure}
       
    \medskip 
    
    \begin{minipage}{\textwidth}
    \footnotesize
    \textbf{Notes:} We first generate the education level as a multinomial variable with four categories (primary school, middle school, high school, and higher education), and then we compute the proportion of fathers (mothers) with the same education level. The lowest proportion possible is 25\% because we consider four categories.
    \end{minipage}
       \caption{Counterfactual experiment on education of parents.}
          \label{fig:simueducation}
\end{figure}

The fourth policy experiment varies the proportion of fathers  or mothers who share the same level of education. In Figure \ref{fig:simueducation}, we vary the proportion of fathers with the same education level (panel (a)) and the proportion of mothers with the same education level (panel (b)). This allows us to observe how changes in education homogeneity  impact network characteristics.

When we increase the proportion of parents with similar education levels this leads to networks with higher degree, density, and clustering. Interestingly, the magnitude of the impact as we move from the status quo to having a perfectly homogeneous composition of parents in terms of education is similar to that found for same-gender children. This finding constitutes another important benchmark, allowing us to distill the importance of children's gender for network structure.

\paragraph{Parents' Age}

\begin{figure}[!htbp]
    \centering
    \begin{subfigure}{\textwidth} 
        \centering
    \includegraphics[scale = 1, width=\textwidth]{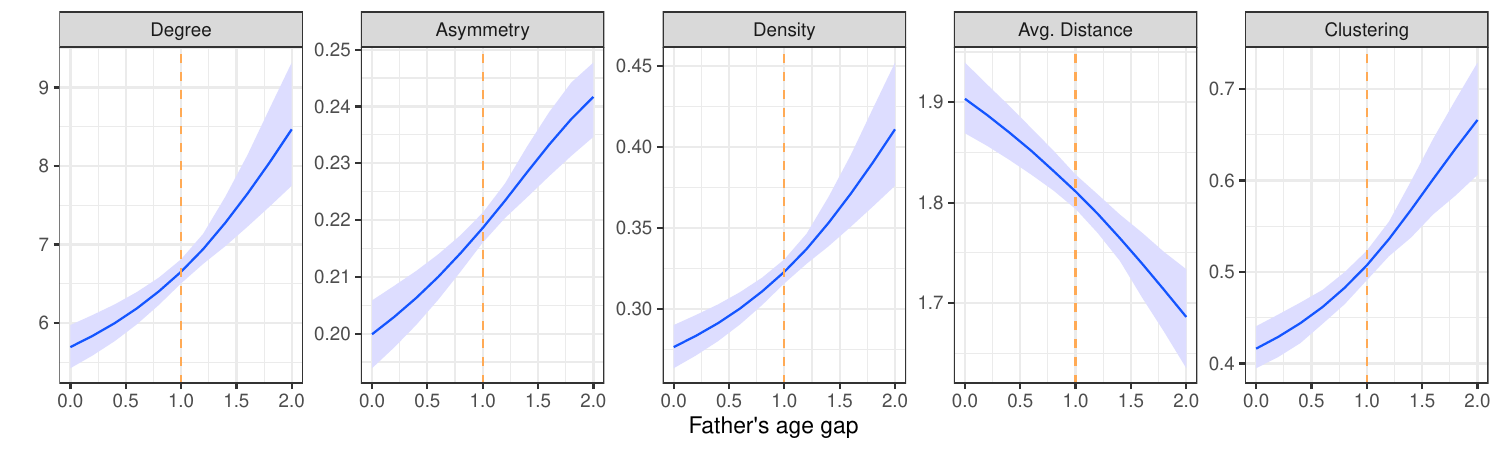}
    \caption{Impact of father's age gap}
     
    \end{subfigure}
    
    \bigskip 
    
    \begin{subfigure}{\textwidth}
        \centering
    \includegraphics[scale = 1, width=\textwidth]{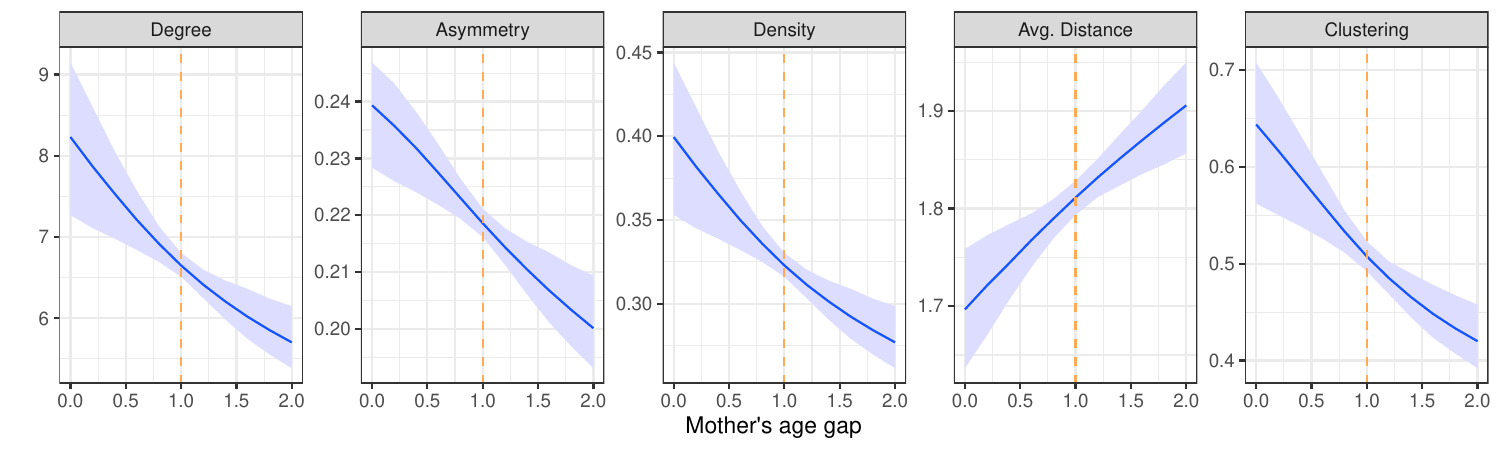}
    \caption{Impact of mother's age gap}
       
    \end{subfigure}
       
    \medskip 
    
    \begin{minipage}{\textwidth}
    \footnotesize
    \textbf{Notes:} For each simulation, we vary the father's (mother's) age gap by regenerating the variable father's (mother's) age as $\hat{\text{Age}} = \text{Age}_0 + \Bar{\delta}\times (\text{Age} - \text{Age}_0)$, where $\text{Age}$ in the observed father's (mother's) age in the original sample, $\text{Inc}_0$ is the average age observed in the original sample, and $\Bar{\delta} $ is the inequality measure. If $\Bar{\delta} = 0$, then the generated age is the observed average age, i.e. the generated age is the same for all families. If $\Bar{\delta} = 1$, then the generated age coincides with the observed age. If $\Bar{\delta} = 2$, then the standard deviation of the generated age is twice that of the observed age.
    
    \end{minipage}
       \caption{Counterfactual experiment on age of parents.}
          \label{fig:simuage}
\end{figure}

In our last experiment, displayed in Figure \ref{fig:simuage} we vary the age of parents to assess its influence on network structure. Panel (a) examines the impact of variations in the age gap among fathers, while 
panel (b) focuses on the age gap among mothers.  

We find some interesting opposing patterns for the two parents. While for fathers, a more dispersed age distribution  increases connections, and enhances the density and clustering of networks, the opposite is true for mothers' age. This suggests that women tend to form connections with other women of the same age while the opposite is true for men. In terms of magnitude, increasing age inequality for fathers  or decreasing that of mothers has a similar impact on network structure as the child's gender.

\section{Conclusion\label{sec:conclusion}}

This paper investigates the role of children's gender on shaping parental network structure. We draw on a rich network dataset of households with children in early childhood comprising 4,500 households across 222 villages in Bangladesh. We estimate a structural model of network formation and simulate it to gain insight as to how the gender composition of children in a village affects the structure of social ties among parents. Our findings reveal that children's gender plays a pivotal role in shaping the number of social connections among parents, with particularly pronounced effects when families have girls, suggesting a nuanced interplay between children's gender and parents' social network formation. Importantly, the impact of children's gender on parental network formation is comparable in magnitude to other key socioeconomic factors, including income distribution, parental occupation, education, and age.

Our findings may have policy implications that favor single-sex schools. However, this needs to be balanced by considering the impact on children's socialization into gender roles, which  begins at an early age. Indeed, in  the psychology literature, it is well-documented that parents begin to socialize their children as boys or girls without even knowing it \citep{Lindsey2011}. Our study shows that parents are more likely to interact with their peers\textemdash here, other parents with children of the same gender\textemdash  which may reinforce gender-typed expectations and stereotypes, so that parents will have a  differential treatment of daughters and sons. Thus, promoting single-sex schools may increase the social interactions of parents and eventually that of children of the same gender but it may also inadvertently promote gender stereotypes.

While our study offers significant insights into parental network formation in a rural Bangladeshi context, future research could extend the analysis to other settings and cultural contexts to enrich our understanding of these dynamics. Additionally, expanding the study to cover families with children from a broader range of age groups could shed light on how these influences on parents' networks evolve throughout children's development. Furthermore, given that parents' social interactions are influenced by their children's gender, a natural next step is to explore the extent to which this channel can explain the formation of social links among children. These are promising avenues that we intent to pursue in future research.

\bibliographystyle{aer}
\bibliography{References}

\newpage 

\setcounter{footnote}{0}

\pagenumbering{arabic}
\renewcommand{\thepage}{A\arabic{page}}

\setcounter{table}{0} \renewcommand{\thetable}{A\arabic{table}}

\setcounter{figure}{0} \renewcommand{\thefigure}{A\arabic{figure}}

\begin{center}

\Large{\textbf{APPENDIX}}
    
\end{center}

\appendix
\section{Estimation method}\label{append:est}

The MCMC approach proposed by \cite{mele2017structural} avoids the need of computing the denominator in the distribution of $A_r$ given by Equation \eqref{eq:pi}. This approach requires simulating at each iteration an artificial network $\tilde A_r$ with the same likelihood than that of the observed network $A_r$. Let $\boldsymbol{\beta} \in \mathbb{R}^{K}$ be the vector of parameters to be estimated, for some integer $K > 0$. With the specification
\begin{align*}
    u(\boldsymbol{x}_{r,i}, \boldsymbol{x}_{r,j}) &= \beta^{u}_{1} Gender_{r,ij} + \boldsymbol{x}^{u\prime}_{ij}\boldsymbol{\beta}^{u}_{2},\\
    m(\boldsymbol{x}_{r,i}, \boldsymbol{x}_{r,j}) &= \beta^{m}_{1} Gender_{r,ij} + \boldsymbol{x}^{m\prime}_{ij}\boldsymbol{\beta}^{m}_{2},\\
    v(\boldsymbol{x}_{r,i}, \boldsymbol{x}_{r,k}) &= \beta^{v}_{1} Gender_{r,ik} + \boldsymbol{x}^{v\prime}_{ik}\boldsymbol{\beta}^{v}_{2},
\end{align*}
we use the notation $Q_r(A_r, \mathcal{X}_r, \boldsymbol{\beta})$ to denote the potential function in Equation \eqref{eq:potential} instead of $Q_r(A_r, \mathcal{X}_r)$.
Given the high number of observations in our sample, we set non-informative flat prior for $\boldsymbol{\beta}$, i.e., the prior density function of $\boldsymbol{\beta}$ is $p(\boldsymbol{\beta}) \propto 1$. We propose an adaptive version of the MCMC process.

\begin{algorithm}
\caption{Simulation from the parameter posterior distribution}\label{alg:post}
\noindent Initialize $\boldsymbol{\beta}^{(0)} \in \mathbb{R}^{K}$ and choose $\tau^{(0)} > 0$ and $\boldsymbol{\Sigma}$ a positive definite matrix dimension $K\times K$.\\
For t from 1 to $T$:
\begin{enumerate}[noitemsep]
    \item  Draw a proposal $\tilde{\boldsymbol{\beta}}$ from $N\left(\boldsymbol{\beta}^{(t - 1)}, \tau^{(t-1)}\boldsymbol{\boldsymbol{\Sigma}}\right)$.
    \item  For $r$ from 1 to $\bar{r} = 222$, simulate the artificial network $\tilde A_r$ using Algorithm \ref{alg:simnet}, where $\boldsymbol{\beta}$ is set to $\tilde{\boldsymbol{\beta}}$ and $\tilde A_r^{(0)}$ is the observed network  $A_r$. We perform this step in parallel for many villages simultaneously because the simulations of $\tilde A_r$ are independent across $r$, conditional on $\boldsymbol{\beta}$.  
    \item Accept the proposal $\tilde{\boldsymbol{\beta}}$ with the probability
    $$\min\left\{1, \dfrac{\exp\left \{\sum_{r = 1}^{\bar r}\left (Q_r(\tilde A_r, \mathcal{X}_r, \boldsymbol{\beta}^{(t-1)}) + Q_r(A_r, \mathcal{X}_r, \tilde{\boldsymbol{\beta}})\right)\right\}}{\exp\left \{\sum_{r = 1}^{\bar r}\left (Q_r(\tilde A_r, \mathcal{X}_r, \tilde{\boldsymbol{\beta}}) + Q_r(A_r, \mathcal{X}_r, \boldsymbol{\beta}^{(t-1)})\right)\right\}}\right\}.$$
    Update $\boldsymbol{\beta}^{(t)} = \tilde{\boldsymbol{\beta}}$ if $\tilde{\boldsymbol{\beta}}$ is accepted, and $\boldsymbol{\beta}^{(t)} = \boldsymbol{\beta}^{(t - 1)}$ otherwise.
    \item Update $\tau^{(t)}$ following \cite{atchade2005adaptive} for the acceptance rate equals 23.4\%.
\end{enumerate}
\end{algorithm}

\noindent We first run Algorithm \ref{alg:post}, defining $\boldsymbol{\Sigma}$ as an identity matrix, and $\tau^{(0)} = 1$. This means that the proposal $\tilde{\boldsymbol{\beta}}$ does not take into account the correlation between the components of $\boldsymbol{\beta}$. We perform $T = 100,000$ simulations and define a new $\boldsymbol{\Sigma}$ as a covariance matrix of the last 50,000 simulations. We run the algorithm a second time with the new $\boldsymbol{\Sigma}$ and perform $T = 100,000$ simulations again. The last 50,000 simulations are considered draws from the posterior distribution of $\boldsymbol{\beta}$. As we rely on an adaptive MCMC in which $\tau^{(t)}$ is updated, we do not care about the choice of $\tau^{(0)}$. We set $\tau^{(0)} = 1$ every time. We do this exercise using several starting values $\boldsymbol{\beta}^{(0)}$ and converge to the same posterior distribution. This suggests that the model does not suffer from the identification problem pointed out by M17.

To simulate the artificial network $\tilde A_r$ in step 2 of Algorithm \ref{alg:post}, we rely on a Gibbs sampler, where only one entry of the network (randomly chosen) is updated at each iteration. We also try to update two and three entries simultaneously at each iteration. We notice that the results are robust to the number of entries we select. The convergence is faster when we select many entries, but the computations become more intensive.

\begin{algorithm}
\caption{Simulation of artificial networks in the village $r$}\label{alg:simnet}
\noindent Inputs: Assign a value to $\boldsymbol{\beta}$ and initialize $\tilde A_r^{(0)}$ is the observed network $A_r$.\\
Perform the following steps for t from 1 to $R = 7,000$.
\begin{enumerate}[noitemsep]
    \item  Randomly chosen the entry $(i, j)$ to update in $\tilde A_r^{(t)}$. We have two proposals for $\tilde A_r^{(t)}$. The $(i, j)$-th entry is zero for the first proposal denoted $A_r^0$ and is one for the second proposal denoted  $A_r^1$. The other entries in $A_r^0$ and $A_r^1$ keep the same values as in $\tilde A_r^{(t - 1)}$.
    \item  Accept $A_r^0$ with the probability 
    $\dfrac{\exp\left \{Q_r(A_r^0, \mathcal{X}_r, \boldsymbol{\beta})\right\}}{\exp\left \{Q_r(A_r^0, \mathcal{X}_r, \boldsymbol{\beta})\right\} + \exp\left \{Q_r(A_r^1, \mathcal{X}_r, \boldsymbol{\beta})\right\}}$.\\
    Set $\tilde A_r^{(t)} = A_r^0$ if $A_r^0$ is accepted and $\tilde A_r^{(t)} = A_r^1$ otherwise.
\end{enumerate}
The last simulation $\tilde A_r^{(R)}$ is the generated artificial network $\tilde A_r$.
\end{algorithm}

\section{Counterfactual analysis}\label{append:countf}
Our counterfactual analysis consists of varying family observable characteristics and examining how these variations influence the network. For example, we can regenerate the variable $Gender$ from a Bernoulli distribution to change the proportion of same-gender families from 50\% to 100\%. Another example is to vary the proportion of boys in the villages from 0\% to 100\%. Each of these examples yields a set of family observable characteristics denoted as $\hat{\mathcal{X}}_r$, which serves as the counterpart of $\mathcal{X}_r$ in our counterfactual experiment. We use $\hat{\mathcal{X}}_r$ to simulate the village networks.

We replicate the following steps 1,000 times.
\begin{enumerate}[noitemsep]
    \item We randomly draw $\boldsymbol{\beta}$ from the last 50,000 simulations of Algorithm \ref{alg:post}.
    \item We simulate $\tilde A_r$ using Algorithm \ref{alg:simnet}, where $\boldsymbol{\beta}$ is set to the simulation of step 1 and $\mathcal{X}_r$ is replaced with $\hat{\mathcal{X}}_r$.
    \item We compute some statistics of $\tilde A_r$, such as degree, clustering, diameter, etc.
\end{enumerate}

\noindent For each statistic, we have 1,000 replications, which can be used to compute the confidence interval.
\section{Supplementary results}
\begin{table}[!htbp]
  \centering 
  \footnotesize
  \addtolength{\leftskip} {-0.5cm}
  \caption{Summary of the parameter posterior distribution (continued)} 
\resizebox{1.1\textwidth}{!}{
\begin{threeparttable}
\begin{tabular}{ld{7}ld{7}ld{7}ld{7}l}
\toprule
                                      & \multicolumn{2}{c}{Model 7} & \multicolumn{2}{c}{Model 8} & \multicolumn{2}{c}{Model 9} & \multicolumn{2}{c}{Model 10} \\ \midrule
\textbf{Direct links}                 & \textbf{}      & \textbf{}  & \textbf{}      & \textbf{}  & \textbf{}      & \textbf{}  & \textbf{}      & \textbf{}  \\
Intercept                             &                                        &            &                                       &            & -1.749^{***} & (0.069)    & -1.734^{***} & (0.073)    \\
$\text{Gender}_{ij}$                  &                                        &            &                                       &            & 0.010^{}     & (0.023)    &                                 &            \\
$\text{Male}_{ij}$                    &                                        &            &                                       &            &                                 &            & 0.007^{}     & (0.026)    \\
$\text{Female}_{ij}$                  &                                        &            &                                       &            &                                 &            & -0.012^{}    & (0.028)    \\
Adiff(Father’s Age)                   &                                        &            &                                       &            & 0.003^{}     & (0.003)    & 0.004^{}     & (0.003)    \\
$\text{Age Father}_i$                 &                                        &            &                                       &            & 0.001^{}     & (0.002)    & 0.001^{}     & (0.002)    \\
$\text{Age Father}_j$                 &                                        &            &                                       &            & -0.005^{**}  & (0.002)    & -0.005^{**}  & (0.003)    \\
Adiff(Mother’s Age)                   &                                        &            &                                       &            & -0.009^{***} & (0.004)    & -0.010^{**}  & (0.004)    \\
$\text{Age Mother}_i$                 &                                        &            &                                       &            & -0.001^{}    & (0.003)    & -0.001^{}    & (0.003)    \\
$\text{Age Mother}_j$                 &                                        &            &                                       &            & 0.007^{**}   & (0.003)    & 0.007^{***}  & (0.003)    \\
Adiff(Income)                         &                                        &            &                                       &            & -0.007^{***} & (0.003)    & -0.007^{***} & (0.002)    \\
$\text{income}_i$                     &                                        &            &                                       &            & 0.008^{***}  & (0.002)    & 0.007^{***}  & (0.002)    \\
$\text{Income}_j$                     &                                        &            &                                       &            & 0.009^{***}  & (0.002)    & 0.010^{***}  & (0.002)    \\
Adiff(Homestead land)                 &                                        &            &                                       &            & 0.004^{}     & (0.113)    & 0.007^{}     & (0.121)    \\
$\text{Homestead land}_i$             &                                        &            &                                       &            & 0.242^{**}   & (0.097)    & 0.246^{**}   & (0.097)    \\
$\text{Homestead land}_j$             &                                        &            &                                       &            & 0.079^{}     & (0.099)    & 0.076^{}     & (0.101)    \\
Adiff(Farming land)                   &                                        &            &                                       &            & -0.121^{***} & (0.015)    & -0.123^{***} & (0.013)    \\
$\text{Farming land}_i$               &                                        &            &                                       &            & 0.120^{***}  & (0.014)    & 0.122^{***}  & (0.013)    \\
$\text{Farming land}_j$               &                                        &            &                                       &            & 0.133^{***}  & (0.014)    & 0.134^{***}  & (0.013)    \\
$\text{Father’s education}_{ij}$      &                                        &            &                                       &            & 0.018^{}     & (0.022)    & 0.015^{}     & (0.026)    \\
$\text{Mother’s education}_{ij}$      &                                        &            &                                       &            & -0.003^{}    & (0.022)    & -0.002^{}    & (0.024)    \\
$\text{Father is daily laborer}_{ij}$ &                                        &            &                                       &            & 0.199^{***}  & (0.033)    & 0.202^{***}  & (0.037)    \\
$\text{Father is farmer}_{ij}$        &                                        &            &                                       &            & 0.053^{}     & (0.047)    & 0.049^{}     & (0.045)    \\
$\text{Father is sel-femployed}_{ij}$ &                                        &            &                                       &            & -0.115^{***} & (0.037)    & -0.112^{***} & (0.035)    \\
$\text{Mother stays home}_{ij}$       &                                        &            &                                       &            & -0.080^{**}  & (0.034)    & -0.081^{***} & (0.034)    \\\midrule
\textbf{Mutual links}                 &                                        &            &                                       &            &                                 &            &                                 &            \\
Intercept                             & -0.258^{***}        & (0.042)    & -0.266^{***}       & (0.044)    & 2.139^{***}  & (0.082)    & 2.109^{***}  & (0.083)    \\
$\text{Gender}_{ij}$                  & 0.026^{}            & (0.023)    &                                       &            & 0.010^{}     & (0.045)    &                                 &            \\
$\text{Male}_{ij}$                    &                                        &            & 0.059^{**}         & (0.027)    &                                 &            & 0.050^{}     & (0.052)    \\
$\text{Female}_{ij}$                  &                                        &            & 0.024^{}           & (0.027)    &                                 &            & 0.047^{}     & (0.057)    \\
Adiff(Father’s Age)                   & -0.001^{}           & (0.003)    & -0.001^{}          & (0.003)    & -0.005^{}    & (0.005)    & -0.005^{}    & (0.005)    \\
Adiff(Mother’s Age)                   & \textless{}0.000^{} & (0.003)    & \textless 0.000^{} & (0.003)    & 0.013^{*}    & (0.007)    & 0.014^{**}   & (0.007)    \\
Adiff(Income)                         & -0.007^{***}        & (0.002)    & -0.007^{***}       & (0.002)    & -0.010^{**}  & (0.005)    & -0.009^{**}  & (0.004)    \\
Adiff(Homestead land)                 & -0.268^{***}        & (0.093)    & -0.261^{***}       & (0.094)    & -0.668^{***} & (0.167)    & -0.676^{***} & (0.169)    \\
Adiff(Farming land)                   & -0.003^{}           & (0.005)    & -0.003^{}          & (0.005)    & -0.016^{*}   & (0.010)    & -0.015^{}    & (0.010)    \\
$\text{Father’s education}_{ij}$      & 0.091^{***}         & (0.024)    & 0.089^{***}        & (0.025)    & 0.042^{}     & (0.045)    & 0.051^{}     & (0.053)    \\
$\text{Mother’s education}_{ij}$      & 0.061^{***}         & (0.024)    & 0.063^{**}         & (0.024)    & 0.055^{}     & (0.044)    & 0.056^{}     & (0.046)    \\
$\text{Father is daily laborer}_{ij}$ & 0.268^{***}         & (0.034)    & 0.266^{***}        & (0.033)    & -0.038^{}    & (0.066)    & -0.046^{}    & (0.070)    \\
$\text{Father is farmer}_{ij}$        & 0.211^{***}         & (0.046)    & 0.206^{***}        & (0.046)    & 0.103^{}     & (0.097)    & 0.112^{}     & (0.089)    \\
$\text{Father is sel-femployed}_{ij}$ & -0.057^{}           & (0.038)    & -0.061^{}          & (0.039)    & 0.109^{}     & (0.074)    & 0.099^{}     & (0.070)    \\
$\text{Mother stays home}_{ij}$       & -0.073^{**}         & (0.033)    & -0.073^{**}        & (0.034)    & 0.071^{}     & (0.066)    & 0.070^{}     & (0.069)  \\\bottomrule
\end{tabular}
\begin{tablenotes}[para,flushleft]
\footnotesize
Note: ***: \textit{pvalue $<1\%$}, **: \textit{pvalue $<5\%$}, *: \textit{pvalue $<10\%$}\\
This table presents the mean of the posterior distribution of each parameter followed by the standard deviation in brackets using the last 50,000 simulations of the MCMC. For any variable denoted "var", "var$_{ij}$" takes one if var$_i =$ var$_j$ and zero otherwise; "Adiff(var)" if the absolute difference between var$_i$ and var$_j$. The network is directed and the relationship we model is the link from $j$ to $i$.
\end{tablenotes}
\end{threeparttable}}
\label{TAB:Post2}
\end{table}

\begin{figure}[htbp]
    \centering
    \includegraphics[scale = 1, width=\textwidth]{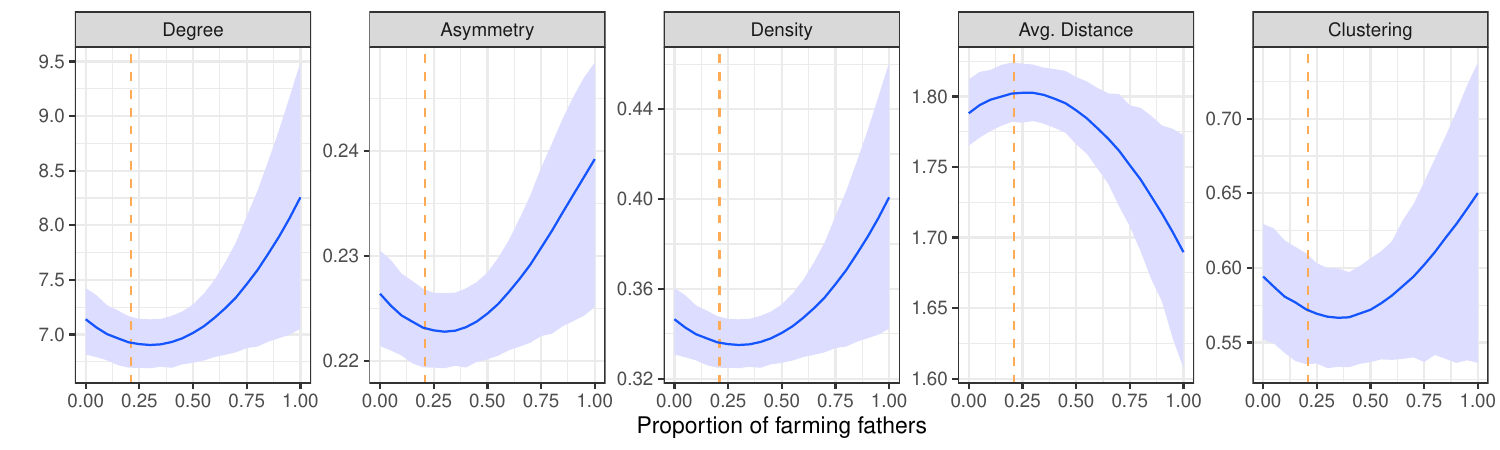}
    \caption{Impact of the proportion of farming fathers}
    \label{fig:simufarmers}
    \justify{\footnotesize Note: Father being a farmer is generated as a dummy variable from a Bernoulli distribution. Changing the variable "farming fathers" also requires changing the proportion of other occupations. Once the variable "farming fathers" is generated, we proportionally adjust the proportion of the other occupations for the total proportion to equal one.}
\end{figure}

\begin{figure}[htbp]
    \centering
    \includegraphics[scale = 1, width=\textwidth]{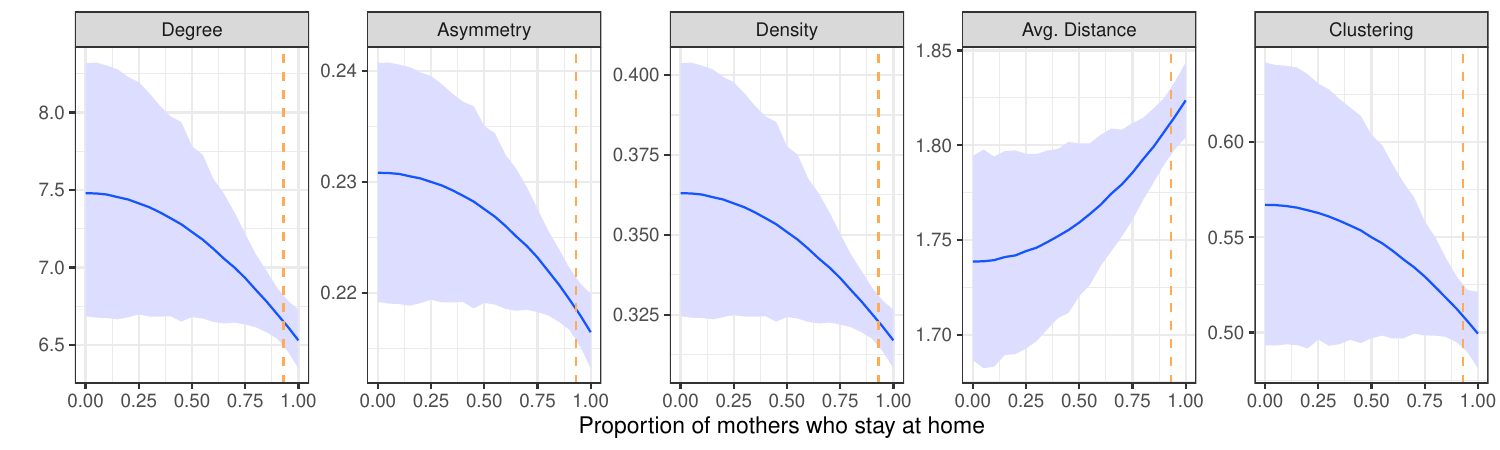}
    \caption{Impact of the proportion of mothers who stay at home}
    \label{fig:simumhome}
    \justify{\footnotesize Note: We vary the proportion of mothers who stay at home by simulating the variable "Mother stays home" from a Bernoulli.}
\end{figure}

\end{document}